\documentclass[twocolumn]{aastex63}

\newcommand{\rpone}{$1.301^{+0.058}_{-0.051}$}
\newcommand{\mpone}{$4.65^{+0.19}_{-0.28}$}
\newcommand{\aone}{$0.0765^{+0.0036}_{-0.0032}$}
\newcommand{\teqone}{$1374^{+34}_{-35}$}
\newcommand{\tcircone}{$18.1^{+2.8}_{-2.3}$}
\newcommand{\rhopone}{$2.61\pm0.38$}
\newcommand{\loggpone}{$3.832^{+0.042}_{-0.052}$}
\newcommand{\faveone}{$0.808^{+0.082}_{-0.078}$}
\newcommand{\msinione}{$4.65^{+0.19}_{-0.28}$}
\newcommand{\periodref}{$5.55149298\pm0.00000051$}
\newcommand{\tcref}{$7147.05248\pm0.00019$}
\newcommand{\ttref}{$7147.05248\pm0.00019$}
\newcommand{\tsref}{$2457144.377^{+0.025}_{-0.024}$}
\newcommand{\tzeroref}{$9184.450409\pm0.000039$}
\newcommand{\iref}{$89.68^{+0.23}_{-0.32}$}
\newcommand{\bref}{$0.059^{+0.059}_{-0.042}$}
\newcommand{\kref}{$448.8\pm7.5$}
\newcommand{\rprsref}{$0.086991^{+0.00011}_{-0.000090}$}
\newcommand{\arref}{$10.70\pm0.17$}
\newcommand{\deltaref}{$0.007567^{+0.000019}_{-0.000016}$}
\newcommand{\deltarref}{$0.00882^{+0.00024}_{-0.00022}$}
\newcommand{\deltazref}{$0.00813^{+0.00017}_{-0.00016}$}
\newcommand{\deltatref}{$0.008718\pm0.000084$}
\newcommand{\tauref}{$0.014432^{+0.00017}_{-0.000049}$}
\newcommand{\tonefourref}{$0.17932^{+0.00016}_{-0.00015}$}
\newcommand{\tfwhmref}{$0.16485\pm0.00014$}
\newcommand{\ruoneref}{$0.286\pm0.045$}
\newcommand{\rutworef}{$0.292\pm0.048$}
\newcommand{\zuoneref}{$0.282^{+0.040}_{-0.041}$}
\newcommand{\zutworef}{$0.322\pm0.046$}
\newcommand{\tuoneref}{$0.140\pm0.037$}
\newcommand{\tutworef}{$0.261\pm0.045$} 

\providecommand{\bjdtdb}{\ensuremath{\rm {BJD_{TDB}}}}

\providecommand{\msun}{\ensuremath{\,M_\Sun}}
\providecommand{\rsun}{\ensuremath{\,R_\Sun}}
\providecommand{\lsun}{\ensuremath{\,L_\Sun}}
\providecommand{\mj}{\ensuremath{\,M_{\rm J}}}
\providecommand{\rj}{\ensuremath{\,R_{\rm J}}}

\providecommand{\fave}{\langle F \rangle}
\providecommand{\fluxcgs}{10$^9$ erg s$^{-1}$ cm$^{-2}$}
\usepackage{apjfonts}
\usepackage{color, colortbl}
\usepackage{multirow}
\usepackage{wrapfig}

\usepackage[first=0,last=9]{lcg}
\definecolor{celadon}{rgb}{0.67, 0.88, 0.69}

\usepackage{amsmath}
\usepackage{amssymb}
\usepackage{hyperref}
\usepackage{afterpage}

\accepted{}

\shorttitle{Reanalysis of KELT-24~\MakeLowercase{b}}
\shortauthors{Giovinazzi et al.}

\begin{document}

\title{Trials and Tribulations in the Reanalysis of KELT-24~b: a Case Study for the Importance of Stellar Modeling}

\author[0000-0002-0078-5288]{Mark R. Giovinazzi}
\affiliation{Department of Physics and Astronomy, University of Pennsylvania, 209 South 33rd Street, Philadelphia, PA 19104 USA}

\correspondingauthor{Mark R. Giovinazzi}
\email{markgio@sas.upenn.edu}

\author[0000-0002-0786-7307]{Bryson Cale}
\affiliation{NASA JPL, 4800 Oak Grove Drive, Pasadena, CA 91109, USA}
\affiliation{IPAC, 770 South Wilson Avenue, Pasadena, CA 91125, USA}

\author[0000-0003-3773-5142]{Jason D. Eastman}
\affiliation{Center for Astrophysics \textbar \ Harvard \& Smithsonian, 60 Garden Street, Cambridge, MA 02138, USA}

\author[0000-0001-8812-0565]{Joseph E. Rodriguez}
\affiliation{Center for Data Intensive and Time Domain Astronomy, Department of Physics and Astronomy, Michigan State University, East Lansing, MI 48824, USA}

\author[0000-0002-6096-1749]{Cullen H. Blake}
\affiliation{Department of Physics and Astronomy, University of Pennsylvania, 209 South 33rd Street, Philadelphia, PA 19104 USA}

\author[0000-0002-3481-9052]{Keivan G. Stassun}
\affiliation{Department of Physics and Astronomy, Vanderbilt University, Nashville, TN 37235, USA}
\affiliation{Department of Physics, Fisk University, 1000 17th Avenue North, Nashville, TN 37208, USA}

\author[0000-0001-7246-5438]{Andrew Vanderburg}
\affiliation{Department of Physics and Kavli Institute for Astrophysics and Space Research, Massachusetts Institute of Technology, Cambridge, MA 02139, USA}

\author[0000-0001-9269-8060]{Michelle Kunimoto}
\affiliation{Department of Physics and Kavli Institute for Astrophysics and Space Research, Massachusetts Institute of Technology, Cambridge, MA 02139, USA}

\author[0000-0001-9811-568X]{Adam L. Kraus}
\affiliation{Department of Astronomy, The University of Texas at Austin, Austin, TX 78712, USA}

\author[0000-0002-6778-7552]{Joseph Twicken}
\affiliation{SETI Institute, Mountain View, CA 94043, USA}
\affiliation{NASA Ames Research Center, Moffett Field, CA 94035, USA}

\author[0000-0002-9539-42039]{Thomas G. Beatty}
\affiliation{Department of Astronomy, 475 N. Charter Street, University of Wisconsin--Madison, Madison, WI 53706, USA}



\author[0000-0001-9408-8848]{Cayla M. Dedrick}
\affiliation{Department of Astronomy \& Astrophysics, 525 Davey Laboratory, The Pennsylvania State University, University Park, PA 16802, USA}
\affiliation{Center for Exoplanets and Habitable Worlds, 525 Davey Laboratory, The Pennsylvania State University, University Park, PA 16802, USA}

\author[0000-0002-1160-7970]{Jonathan Horner}
\affiliation{Centre for Astrophysics, University of Southern Queensland, USQ Toowoomba, QLD 4350, Australia}

\author[0000-0002-1159-1083]{John A. Johnson}
\affiliation{Center for Astrophysics \textbar \ Harvard \& Smithsonian, 60 Garden Street, Cambridge, MA 02138, USA}

\author[0000-0001-9397-4768]{Samson A. Johnson}
\affiliation{NASA JPL, 4800 Oak Grove Drive, Pasadena, CA 91109, USA} 

\author[0000-0002-8041-1832]{Nate McCrady}
\affiliation{Department of Physics and Astronomy, University of Montana, 32 Campus Drive, No. 1080, Missoula, MT 59812, USA}

\author[0000-0002-8864-1667]{Peter Plavchan}
\affiliation{George Mason University, 4400 University Drive, Fairfax, VA 22030, USA}

\author[0000-0002-6228-8244]{David H. Sliski}
\affiliation{David R. Mittelman Observatory, Mayhill, NM 88339, USA}

\author[0000-0003-1928-0578]{Maurice L. Wilson}
\affiliation{High Altitude Observatory, National Center for Atmospheric Research, 3080 Center Green Drive, Boulder, CO 80301, USA}

\author[0000-0001-9957-9304]{Robert A. Wittenmyer}
\affiliation{Centre for Astrophysics, University of Southern Queensland, USQ Toowoomba, QLD 4350, Australia}

\author[0000-0001-6160-5888]{Jason T. Wright}
\affiliation{Department of Astronomy \& Astrophysics, 525 Davey Laboratory, The Pennsylvania State University, University Park, PA 16802, USA}
\affiliation{Center for Exoplanets and Habitable Worlds, 525 Davey Laboratory, The Pennsylvania State University, University Park, PA 16802, USA}

\author[0000-0002-5099-8185]{Marshall C. Johnson} 
\affiliation{Department of Astronomy, The Ohio State University, 4055 McPherson Laboratory, 140 West 18th Avenue, Columbus, OH 43210, USA}

\author[0000-0003-4724-745X]{Mark E. Rose}
\affiliation{NASA Ames Research Center, Moffett Field, CA 94035, USA}


\author[0000-0003-1012-4771]{Matthew Cornachione}
\affiliation{Oregon Institute of Technology, 3201 Campus Drive, Klamath Falls, OR 97601, USA}


\begin{abstract}

We present a new analysis of the KELT-24 system, comprising a well-aligned hot Jupiter, KELT-24~b, and a bright ($V=8.3$), nearby ($d=96.9~\mathrm{pc}$) F-type host star. KELT-24~b was independently discovered by two groups in 2019, with each reporting best-fit stellar parameters that were notably inconsistent. Here, we present three independent analyses of the KELT-24 system, each incorporating a broad range of photometric and spectroscopic data, including eight sectors of TESS photometry and more than 200 new radial velocities (RVs) from MINERVA. Two of these analyses use KELT-24's observed spectral energy distribution (SED) through a direct comparison to stellar evolutionary models, while our third analysis assumes an unknown additional body contributing to the observed broadband photometry and excludes the SED. Ultimately, we find that the models that include the SED are a poor fit to the available data, so we adopt the system parameters derived without it. We also highlight a single transit-like event observed by TESS, deemed likely to be an eclipsing binary bound to KELT-24, that will require follow-up observations to confirm. We discuss the potential of these additional bodies in the KELT-24 system as a possible explanation for the discrepancies between the results of the different modeling approaches, and explore the system for longer-period planets that may be weakly evident in the RV observations. The comprehensive investigations that we present not only increase the fidelity of our understanding of the KELT-24 system, but also serve as a blueprint for future stellar modeling in global analyses of exoplanet systems.

\end{abstract}


\vspace{0.6cm}

\section{Introduction} \label{sec:intro}

The characterization of exoplanets via modern detection techniques, namely the transit and radial velocity (RV) methods, relies heavily on our foundational understanding of the stars they orbit \citep[e.g.,][]{2018haex.bookE.153M}. For example, quasi-periodic signals from stars, like rotation, spots, or activity cycles, can mimic the effects of substellar companions \citep[e.g.,][and references therein]{2014ApJ...796..132D}, while misunderstanding stellar parameters can impede an accurate classification of confirmed exoplanets. Just days apart, \cite{rodriguez2019} and \cite{2019A&A...631A..76H}, hereafter R2019 and H2019 for concision, independently published the discovery of a transiting planetary companion to HD~93148, though their individual sets of best-fit stellar and planetary parameters were inconsistent. Specifically, their derived stellar masses are in particularly strong ($\sim2.4\sigma$) disagreement; R2019 found $M_*=1.460^{+0.055}_{-0.059}~\mathrm{M_\odot}$, while H2019 found $M_*=1.30^{+0.04}_{-0.03}~\mathrm{M_\odot}$. In either case, the star is too massive for calibrated photometric mass relationships \citep[e.g.,][]{Mann_MK, Gio_gorp}. This apparent disparity in stellar mass, as well as that of the two reported RV semi-amplitudes ($K=462^{+16}_{-15}~\mathrm{m~s^{-1}}$ from R2019 and $K=415\pm13~\mathrm{m~s^{-1}}$ from H2019), propagated into a near-5$\sigma$ discrepancy in planetary mass; R2019 found $M_\mathrm{p}=5.18^{+0.21}_{-0.22}~\mathrm{M_\mathrm{J}}$, while H2019 found $M_\mathrm{p}=4.20\pm0.20~\mathrm{M_\mathrm{J}}$. Since both groups observed multiple transits, consensus was reached on a 5.55~d orbital period; R2019 and H2019 used photometry from the Kilodegree Extremely Little Telescope (KELT; \citealt{2007PASP..119..923P, 2012PASP..124..230P, 2018haex.bookE.128P}) and the Multi-site All-Sky CAmeRA (MASCARA; \citealt{2017A&A...601A..11T}) surveys, respectively. Hereafter, we follow R2019 and the NASA Exoplanet Archive\footnote{\url{https://exoplanetarchive.ipac.caltech.edu}} in adopting KELT-24 as the system's name (H2019 published the system as MASCARA-3). While the two aforementioned discovery papers produce different interpretations of KELT-24~b, both publications agree that it is a super-Jupiter-mass planet on a prograde, well-aligned, short-period orbit around a bright ($V=8.3$), F-type rapid rotator ($v\sin i\sim$ 20~km~s$^{-1}$).




Hot Jupiters ($M_\mathrm{p}>0.1~\mathrm{M_J}$, $P<10$~d, as we adopt from \citealt{2012ApJ...753..160W}) make up more than 10\% of the $5000+$ exoplanets known today. Their apparent abundance, however, is the result of an observational bias, as hot Jupiters' large masses and close proximity to their host stars make them relatively easy for exoplanet missions to detect. In general, hot Jupiters are estimated to have an occurrence rate of $\sim1\%$ \citep{2012ApJ...753..160W, 2022MNRAS.516...75B}, though their frequency has been shown to be higher for more massive, FGK-type stars when compared to smaller, M-type stars \citep{2010PASP..122..905J, 2013A&A...549A.109B}. The discovery of 51 Pegasi b \citep{1995Natur.378..355M} marked not only the first exoplanet discovered around a solar-type star, but also the first known hot Jupiter, which paved the way for future exoplanet searches around main sequence stars and challenged previous planet-formation theories. 

It is believed that hot Jupiters do not form close to their stars, but rather migrate inwards during the system's evolution. While this transition is not well-understood, there are several leading theories for the origins of hot Jupiters (see \citealt{2018ARA&A..56..175D} and \citealt{2021JGRE..12606629F} for more detailed reviews on hot Jupiter formation hypotheses). One theory is that the orbit of hot Jupiters within the circumstellar gas-dust disk shrinks gradually during the formation of the solar system, perhaps even circularizing \citep{2003ApJ...586..540D}. Additionally, recent work has shown that many hot Jupiters reside in highly eccentric orbits, consistent with the theory of high-eccentricity migration \citep[see, for example, ][]{2021AJ....161..194R, 2023MNRAS.521.2765R, 2022AJ....164...70Y, 2023ApJS..265....1Y}. In any case, a relatively significant number of hot Jupiters are found to be in misaligned orbits with respect to the rotation of their star \citep[e.g.,][]{2010koa..prop..121W, 2012ApJ...757...18A, 2022BAAS...54e1104R}, an indicator that they may result from large-scale perturbation during the initial formation of the circumstellar disk. Presently, fewer than 50 hot Jupiters have measured spin-orbit angles, or projected obliquities, affirming the importance of constraining systems like KELT-24 for which these types of observations are feasible.

Transiting hot Jupiters are amenable for determining the relative alignment between the planet's orbital angular momentum and the star's rotation axis, primarily via the Rossiter-Mclaughlin (RM) effect, which describes the asymmetry that is present in RV measurements during the transit event of an exoplanet \citep{1924ApJ....60...15R, 1924ApJ....60...22M}. Observing the RM effect in exoplanet systems has only become possible in recent years \citep[e.g.,][]{2007ApJ...655..550G, 2016A&A...588A.127C}. A more robust technique for measuring the system's projected obliquity, known as Doppler tomography (DT; \citealt{2010MNRAS.403..151C}), is executed by making high-cadence observations of a star's spectrum during a planetary transit. From ingress to egress, the planet will induce variations in the spectroscopic line profile in the form of a Doppler "shadow" capable of extracting independent determinations of the relative spin-orbit angle and projected stellar rotational velocity. Both RM and DT measurements offer insight into the system's architecture and dynamical history, which is useful for assessing planetary formation theories. Though rapidly rotating ($v \sin i \gtrsim  10~\mathrm{km~s^{-1}}$) stars like KELT-24 are often avoided by exoplanet surveys, particularly by those using Doppler spectroscopy, large rotational velocities result in a more pronounced signal for spin-orbit alignment studies \citep{2013A&A...550A..53B}. While R2019 and H2019 found discrepant solutions in their joint analyses of the KELT-24 system, both confirmed that the orbit of KELT-24~b is well-aligned with the rotation of its star, making it not only one of the most massive planets known to have such a configuration, but also an intriguing case study to test formation theories for short-period, giant planets.

While the independent analyses led in parallel by R2019 and H2019 validated the existence of a hot Jupiter companion to KELT-24, their individual sets of best-fit stellar and planetary parameters were inconsistent. In this work, we attempt to bridge these discrepancies by combining new measurements with all available data and considering a suite of three different global fits to the system. In Section \ref{sec:data}, we summarize the collective photometric and spectroscopic observations used in this analysis. We outline the observed stellar properties of KELT-24 in Section \ref{sec:star}, and describe the various models generated to characterize the KELT-24 system in Section \ref{sec:global_fit}. In Section \ref{sec:other_planets}, we highlight a stellar companion to KELT-24, as well as a single transit observed via photometry, and present them as possible sources of confusion in fitting the system while searching for longer-period planets. We discuss the newly-refined KELT-24 system within the landscape of known hot Jupiters and exoplanets more broadly in Section \ref{sec:discussion}. A summary of our work is provided in Section \ref{sec:conclusion}.


\vspace{-0.09cm}
\section{Observations} \label{sec:data}

KELT-24 was independently reported to have a planet by R2019 and H2019 with ephemerides derived from KELT and MASCARA photometry, respectively. The KELT survey was designed to scan over 85\% of the sky with 20-30 minute cadence with the goal of discovering transiting hot Jupiters around bright ($7<V<12$) stars \citep{2007PASP..119..923P, 2012PASP..124..230P, 2018haex.bookE.128P}. The survey, which collected first light in 2008, has recently been retired due to the launch of next-generation missions like the Transiting Exoplanet Survey Satellite (TESS). KELT used two 42~mm telescopes, one being KELT-North (Winer Observatory outside of Sonoita, Arizona) and the other KELT-South (South African Astronomical Observatory in Sutherland, South Africa). Both sites were equipped with a Mamiya 645 80~mm $f$/1.9 lens, a 4k$\times$4k Apogee CCD with a 9~$\mu$m pixel size (23$\arcsec$/pixel), and a Paramount ME mount. KELT photometry contributed to the discovery of 26 exoplanets, most of which have orbital periods less than eight days. KELT-24 was observed 10,181 times between 2013 Sep 24 and 2017 Dec 31 across two KELT-North (KN) fields KN26 and KN27. MASCARA was also designed to survey a majority of the sky at high cadence ($\lesssim~8$~minute)for transiting exoplanet systems around bright stars \citep{2017A&A...601A..11T}. It received first light in 2016, and is made up of two instrument sites (La Palma, Spain and La Silla, Chile), each housed with five Canon 24~mm $f$/1.4 USM L II lenses with 17~mm apertures and 4k$\times$2.7k CCD with a 9~$\mu$m pixel size. MASCARA has contributed to numerous exoplanet detections, including 27,247 observations of KELT-24/MASCARA-3 between 2015 February and 2018 March published in H2019.

The discovery of KELT-24~b preceded the first data release of TESS. Since the photometric precision of TESS vastly outperforms that of both KELT and MASCARA, original data from those surveys are excluded from our analysis. However, we do include a prior on the transit time derived from KELT photometry to capitalize on the long baseline of observations between KELT and TESS and improve the precision of the ephemeris for future studies. We describe the observations used to reproduce and refine the KELT-24 system in the sub-sections that follow.


\subsection{KELT-FUN Photometry}

The KELT Follow-Up Network (KELT-FUN) is a global network of observatories and independent astronomers with the primary goal of confirming and characterizing transiting exoplanets \citep{2019yCat..51560234C}. R2019 presented six transits of KELT-24~b from seven different observatories across multiple bandpasses using KELT-FUN. Light curves from these data are shown in Figure \ref{fig:all_transits}. All KELT-FUN observations were scheduled using \texttt{TAPIR} \citep{2013ascl.soft06007J} and were reduced using \texttt{AstroImageJ} \citep{2017AJ....153...77C}.

\subsection{TESS Photometry} \label{subsec:tess}

TESS was originally designed for a two-year mission to scan the entire sky \citep{TESS_mission}, though is now entering its sixth year of operation. TESS uses four identical cameras, each with four 2k$\times$2k CCDs (21\arcsec/pixel). Since its launch in 2018, TESS has discovered more than 400 planets, with $\approx5000$ candidates awaiting confirmation. TESS observed KELT-24 (TIC 349827430) with two-minute cadence in eight sectors, as detailed in Table \ref{tab:tess_sec}. The complete dataset of all sectors used in this analysis can be found at MAST: \dataset[10.17909/4agy-rx54]{http://dx.doi.org/10.17909/4agy-rx54}.

\startlongtable
\begin{deluxetable}{ccc}
\tablecaption{TESS Sectors that observed KELT-24}
\tablehead{\colhead{Sector} & \colhead{Dates} & \colhead{\# Transits}}
\startdata
14 & ~~~2019 Jul 18 - 2019 Aug 14 & 5 \\
20 & 2019 Dec 24 - 2020 Jan 20 & 5 \\
21 & ~~2020 Jan 21 - 2020 Feb 18 & 4 \\
40 & 2021 Jun 24 - 2021 Jul 23 & 5 \\
41 & ~~~2021 Jul 23 - 2021 Aug 20 & 5 \\
47 & 2021 Dec 30 - 2022 Jan 28 & ~~4$^*$ \\
48 & ~~2022 Jan 28 - 2022 Feb 26 & 4 \\
60 & 2022 Dec 23 - 2023 Jan 18 & 4 \\
\enddata
\tablenotetext{}{$^*$A 5th transit was removed due to being clipped during the sector's downlink period.}
\label{tab:tess_sec}
\end{deluxetable}

\begin{figure}[h]
    \centering
    \includegraphics[width=\linewidth]{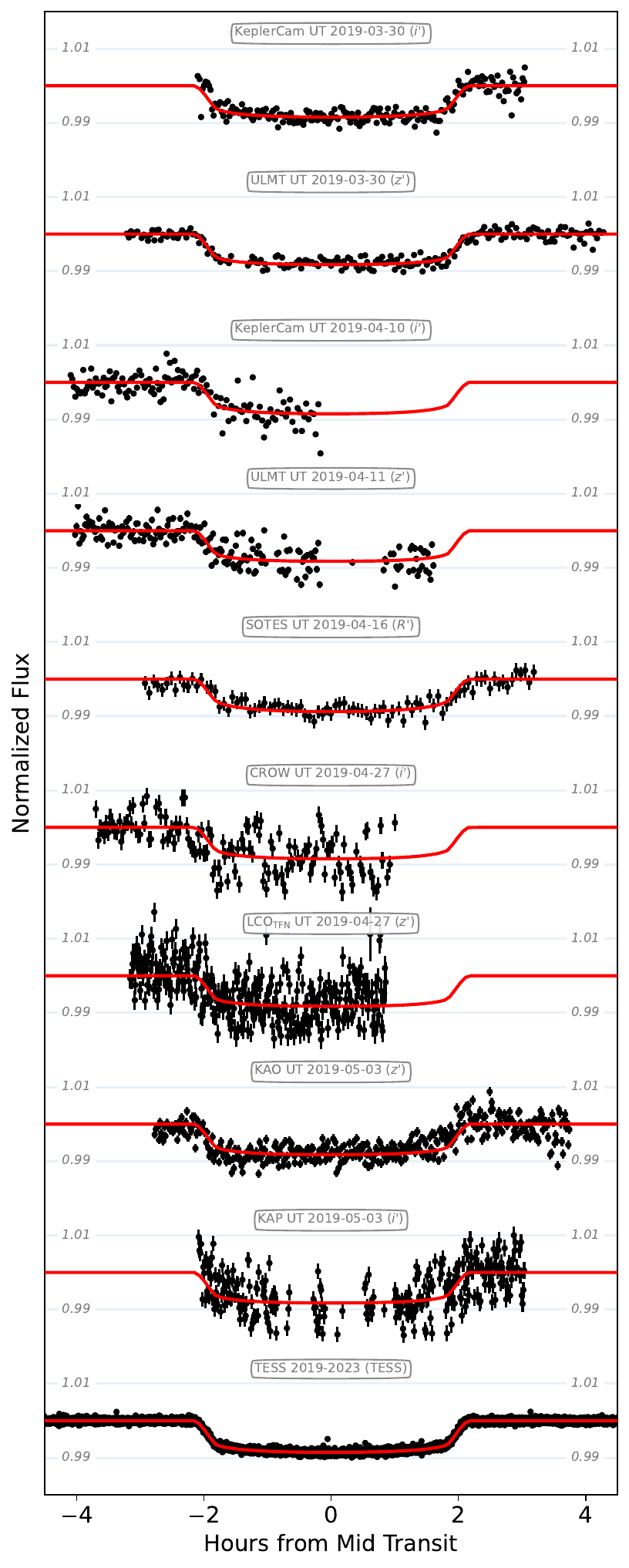}
    \caption{Phase-folded light curves from all KELT-FUN and TESS photometry. Spurious outlier points that deviate more than 2\% from the baseline flux are removed here for enhanced clarity.} 
    \label{fig:all_transits}
\end{figure}

We download all light curves using \texttt{Lightkurve} \citep{2018ascl.soft12013L}, and use the Science Processing Operations Center \citep[SPOC;][]{2016SPIE.9913E..3EJ} ``Presearch Data Conditioning'' light curves utilizing ``Simple Aperture Photometry'' \citep[PDCSAP;][]{2012PASP..124..985S, 2014PASP..126..100S, 2012PASP..124.1000S} in our joint-modeling in Section \ref{sec:global_fit}. Additionally, all light curves were flattened using \texttt{keplerspline} \citep{2014PASP..126..948V}. The phase-folded light curve compiled from the 36 full transits of KELT-24~b captured by TESS are shown in Figure \ref{fig:all_transits}. Transit parameters determined from all photometry used in this analysis are given in Table \ref{tab:transitpars}. 



\startlongtable
\begin{deluxetable}{cccc}
\tablecaption{Median values and 68$\%$ confidence interval for transit times and depths determined across all photometry used in this analysis. The TESS transit time is fit to an intermediate transit between sectors. The TESS depth is computed using the combined phase-folded photometry.}
\tablehead{\colhead{Transit} & \colhead{Band} & \colhead{$T_\mathrm{T}~\left[\mathrm{BJD - 2450000}\right]$} & \colhead{$\mathrm{Depth}~\left[\mathrm{ppm}\right]$}}
\startdata
KeplerCam  & $i'$   & $8573.786124 \pm 0.000070$            & $8717^{+100}_{-100}$\\
ULMT       & $z'$   & $8573.786124 \pm 0.000070$            & $8416^{+72}_{-71}$\\
KeplerCam  & $i'$   & $8584.889110^{+0.000070}_{-0.000069}$ & $8717^{+100}_{-100}$\\
ULMT       & $z'$   & $8584.889110^{+0.000070}_{-0.000069}$ & $8416^{+72}_{-71}$\\
SOTES      & $R$    & $8590.440603 \pm 0.000069$            & $8790^{+170}_{-160}$ \\
CROW       & $i'$   & $8601.543589 \pm 0.000068$            & $8717^{+100}_{-100}$\\
LCO$_\mathrm{TFN}$& $z'$   & $8601.543589 \pm 0.000068$            & $8416^{+72}_{-71}$\\
KAO        & $z'$   & $8607.095082 \pm 0.000068$            & $8416^{+72}_{-71}$\\
KAP        & $z'$   & $8607.095082 \pm 0.000068$            & $8416^{+72}_{-71}$\\
TESS       & TESS & $9162.244385 \pm 0.000039$            & $8534 \pm 17$\\
\enddata
\label{tab:transitpars}
\tablenotetext{}{See \cite{2018AJ....156..234C} for a complete description of all KELT-FUN facilities.}
\end{deluxetable}



\subsection{TRES Spectroscopy} \label{sec:tres_rvs}

R2019 obtained 59 spectra (with a resolving power of $R\sim$~44,000) of KELT-24 from the Tillinghast Reflector Echelle Spectrograph (TRES; \citealt{furesz2008, 2011ASPC..442..305M}) at the Fred Lawrence Whipple Observatory (FLWO) on Mt. Hopkins in Arizona. RV estimates were extracted from the reduced spectra according to the process outlined in \cite{2010ApJ...720.1118B} and \cite{2012ApJ...756L..33Q}. 

Of these 59 spectra, 40 were taken during and immediately after a transit event of KELT-24 the night of 2019 March 31. The DT signal for KELT-24~b was extracted from the in-transit TRES RVs following \cite{2016MNRAS.460.3376Z}, and is included concurrent with the overall fit described in Section \ref{sec:global_fit}. The remaining 19 out-of-transit RVs from TRES were used to constrain the orbit of KELT-24~b.


Here, we add one additional out-of-transit spectrum from TRES taken on 2022 June 15, which extends the RV-baseline to more than four years.

\subsection{SONG Spectroscopy} \label{sec:song_rvs}

Between 2018 April and 2019 May, the Hertzsprung Stellar Observation Network Group (SONG; \citealt{2014RMxAC..45...83A, 2019PASP..131d5003F}) telescope collected 92 RV measurements of KELT-24, 24 of which were taken with high cadence during a transit on 2018 May 29 using a Thorium-Argon (ThAr) lamp for wavelength calibration to determine KELT-24 b’s relative spin-orbit angle. KELT-24~b's relative spin-orbit angle. We do not include the ThAr RVs from SONG. However, we do incorporate a subset of 63 of the remaining published SONG RVs into our updated analysis, two of which are found to have incidentally been taken during a transit. These spurious in-transit RVs are used in conjunction with our DT-fitting to constrain the relative spin-orbit angle of KELT-24~b.


\subsection{MINERVA Spectroscopy} \label{sec:minerva_rvs}

R2019 published 37 RV measurements of KELT-24 from the MINiature Exoplanet Radial Velocity Array (MINERVA; \citealt{swift2015, wilson2019}), all taken on 2019 March 31 to measure KELT-24~b's DT signal. The MINERVA array is a set of four PlaneWave CDK700 0.7~m telescopes at the FLWO on Mt. Hopkins in Arizona. Each exposure consists of spectra from all four telescopes. Of these observations, 17 exposures were collected in-transit and used to supplement the DT analysis outlined in Section \ref{sec:tres_rvs}. The DT signal extraction from the MINERVA in-transit RVs was performed following \cite{2016MNRAS.460.3376Z}. We maintain the simultaneous fitting of the DT signal observed with MINERVA in our updated analysis.


For the purpose of updating the KELT-24 system, we collected 201 new RV measurements from MINERVA between 2019 March 26 and 2020 February 2. Of these, nine are found to have incidentally been taken during transit and are used to help further constrain KELT-24~b's projected obliquity. We publish all RVs to refine the orbit of KELT-24~b and attempt to place constraints on other potential planets in the system. A sample of RVs from our presented MINERVA campaign are provided in Table \ref{tab:minerva_rvs}. All raw MINERVA spectra used in this analysis are reduced and optimally extracted with methods described in \cite{wilson2019}. The corresponding MINERVA RVs are generated using updated methods to those outlined in \cite{caleetal2019}, now implemented in the package \texttt{Echelle.jl}\footnote{\url{https://github.com/astrobc1/Echelle.jl}}. Unlike \cite{wilson2019} where orders are broken into 128-pixel wide chunks, whole echelle orders are modeled at once.

To model the stellar spectrum, we start with a synthetic BT-Settl template \citep{2012RSPTA.370.2765A} corresponding to the parameters of KELT-24 provided in R2019. The template is further Doppler broadened with the \texttt{rotBroad} routine from PyAstronomy \citep{pya} with $v \sin i = 20~\mathrm{km~s^{-1}}$. We provide a summary of all RVs used here in Table \ref{tab:all_rvs}.

\startlongtable
\begin{deluxetable}{cccc}
\tablecaption{Relative RVs for KELT-24 taken from MINERVA}
\tablehead{\colhead{BJD$_\mathrm{TDB}$} & \colhead{RV [m~s$^{-1}$]} & \colhead{$\sigma_\mathrm{RV}$ [m~s$^{-1}$]} & \colhead{Instrument}}
\startdata
2458568.83722 & -269.58 & 116.65 & T1 \\
2458570.77767 & -67.07 & 65.84 & T1 \\ 
2458571.86381 & 293.79 & 65.61 & T1 \\ 
2458574.74811 & -271.90 & 218.97 & T1 \\ 
2458575.79528 & -234.70 & 68.03 & T1 \\ 
... & ... & ... & ... \\ 
2458568.70843 & -326.83 & 85.14 & T2 \\ 
2458570.77767 & -151.08 & 68.97 & T2 \\ 
2458571.86381 & 363.17 & 55.55 & T2 \\ 
2458574.74811 & -465.54 & 120.71 & T2 \\ 
2458575.85455 & -536.59 & 109.12 & T2 \\ 
... & ... & ... & ... \\ 
2458568.70843 & -332.36 & 89.43 & T3 \\ 
2458571.73485 & 426.11 & 81.57 & T3 \\ 
2458574.89035 & -536.88 & 195.3 & T3 \\ 
2458575.85455 & -398.95 & 99.10 & T3 \\ 
2458576.85601 & 79.90 & 70.70 & T3 \\ 
... & ... & ... & ... \\ 
2458570.77767 & -227.12 & 90.01 & T4 \\ 
2458571.92829 & 40.87 & 68.61 & T4 \\ 
2458574.81923 & -225.91 & 125.59 & T4 \\ 
2458575.85455 & -546.18 & 79.88 & T4 \\ 
2458576.85601 & -61.49 & 58.15 & T4 \\ 
... & ... & ... & ... \\ 
\enddata
\label{tab:minerva_rvs}
\tablenotetext{}{T\# denotes the telescope number for MINERVA. The complete table is provided in a machine-readable format.}
\end{deluxetable}

\startlongtable
\begin{deluxetable}{cccc}
\tablecaption{Summary of all RVs for KELT-24}
\tablehead{\colhead{Instrument} & \colhead{$N_\mathrm{obs}$} & \colhead{Dates} & \colhead{$\bar{\sigma_\mathrm{RV}}$}}
\startdata
TRES & 25 & 2019 Mar 06 - 2022 Jun 15 & 32.90 \\
SONG & 65 & 2018 Apr 14 - 2019 Jun 09 & 40.00 \\ 
MINERVA$_\mathrm{T1}$ & 63 & 2019 Mar 26 - 2020 Feb 2 & 125.13 \\
MINERVA$_\mathrm{T2}$ & 47 & 2019 Mar 26 - 2020 Jan 11 & 108.02 \\
MINERVA$_\mathrm{T3}$ & 54 & 2019 Mar 26 - 2019 Nov 24 & 110.90 \\
MINERVA$_\mathrm{T4}$ & 37 & 2019 Mar 28 - 2020 Jan 11 & 148.83 \\
\enddata
\label{tab:all_rvs}
\tablenotetext{}{$\bar{\sigma_\mathrm{RV}}$ is the average RV error for each instrument, in m~s$\mathrm{^{-1}}$.}
\end{deluxetable}

\subsection{Keck/NIRC2 Imaging} \label{sec:imaging}

An epoch of adaptive optics (AO) imaging from the Near Infrared Camera 2 (NIRC2) on the W. M. Keck Observatory was obtained by R2019 on 2019 May 12 in the Br$\gamma$ passband to help account for any extraneous stellar sources that may be contaminating the data taken on KELT-24. They reported a nearby star, with $\Delta$Br$\gamma=2.6$~mag. Here, we reprocess that same image. We also publish an AO observation of the KELT-24 system taken on 2019 June 10 using standard imaging and coronagraphic imaging, as well as non-redundant masking, which has the benefit of achieving higher angular resolution. This allows us to not only extract relative astrometry of the two sources at two independent epochs, but also place more stringent quantitative limits on possible background or unresolved companions that may have previously gone undetected. Details of these observations are presented in Table \ref{tab:keckao}, and contrast limits on hierarchical companions derived from these images are discussed in Section \ref{sec:other_planets}. For the newly published AO measurement, we obtained images in NIRC2's $K_\mathrm{cont}$- and $K_\mathrm{p}$-bands, as well as a coronagraphic observation with the $K_\mathrm{p}$ filter. The image taken with the broader $K_\mathrm{p}$-band, while capable of making deeper photometric observations, was saturated due to the brightness of KELT-24. We applied a remedy to mitigate saturation by flagging and ignoring saturated pixels in PSF-fitting photometry, which maintains the quality of the image for the purpose of searching for fainter sources, but diminishes the quality of the relative astrometry that can be extracted. For this reason, our $K_\mathrm{p}$-band images are useful for placing the contrast detection limits described in Section \ref{sec:other_planets}, while the narrower $K_\mathrm{cont}$-band image is amenable for extracting relative astrometry at the level of the 2019 May 12 $\mathrm{Br}\gamma$ image, providing two independent relative astrometric measures of the pair that remain consistent with our assumption that the pair is bound.

\startlongtable
\begin{deluxetable}{lcccc}
\tablecaption{NIRC2 measurements of KELT-24 and its companion}
\tablehead{\colhead{Epoch} & \colhead{Band} & \colhead{$\rho$ ['']} & \colhead{PA [$^\circ$]} & \colhead{$\Delta m$ [mag]}}
\startdata
2019.3596 &  Br$\gamma$ &  2.07832 &  172.517 &  2.551 $\pm$0.012 \\
2019.4389 &  $K_\mathrm{cont}$       &  2.07806 &  172.524 &  2.520 $\pm$0.020 \\
2019.4389 &  $K_\mathrm{p}$  & -- &  -- &  2.469 $\pm$0.100
\enddata
\tablenotetext{}{Standard errors to the separation $\rho$ and position angle PA are 1.58~mas and 0.040$^\circ$, respectively. Relative astrometry for the $K_\mathrm{p}$-band is not published because the image was saturated, and the associated uncertainties are therefore much larger than those taken in the $K_\mathrm{cont}$-band at the same epoch.}
\label{tab:keckao}
\end{deluxetable}

\subsection{Gaia Astrometry} \label{subsec:gaia}

Gaia is a state-of-the-art space mission with the ultimate goal of creating a precise six-dimensional map of stars in the Milky Way \citep{2016A&A...595A...1G}. The two stars identified here (KELT-24 and the neighboring fainter component) are each catalogued by Gaia DR3 \citep{2023A&A...674A...1G}, with published astrometry averaged at the $t=2016.0$ epoch, and are found to have $\Delta G=5.05$~mag with a corresponding projected separation of $2.06730\pm0.00009\arcsec$ ($s\approx200$~au in the sky-plane projection), assuming an inflation factor of 1.37 in the positional errors following \cite{2021ApJS..254...42B}.

\begin{figure}[h]
    \centering
    \includegraphics[width=\linewidth]{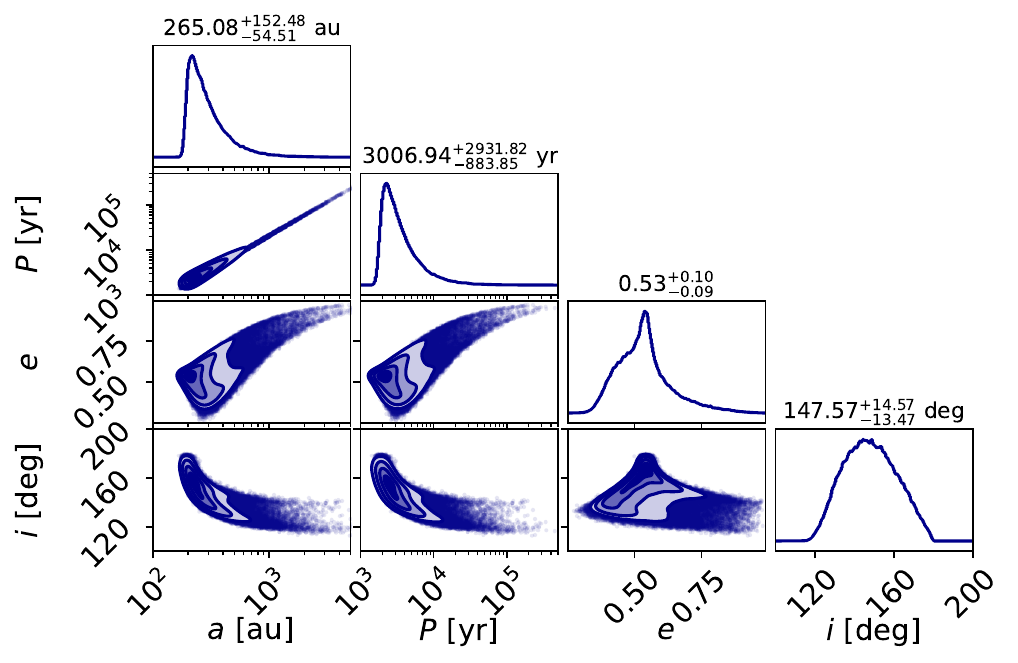}
    \caption{Corner plot from LOFTI accepted trial orbits.}
    \label{fig:KELT24B}
\end{figure}

Their parallaxes and proper motions are also broadly consistent with one another (see Table \ref{tab:observed_stellar_params}), an indicator that they may be bound. \cite{2021MNRAS.506.2269E} searched the Gaia eDR3 catalog \citep{2021A&A...649A...1G}, for binary star systems and determined that KELT-24 and the nearby background star have a chance alignment probability of $\mathcal{R}=5.19\times10^{-6}$, an indicator that there is a very low probability that the two sources are a chance alignment instead of a gravitationally bound system. Under the assumption that these two stars are indeed bound, we simulated 10 million orbits randomly in phase, uniformly in $\cos i$, and thermally ($f\left(e\right)=2e$) in eccentricity to determine that their semimajor axis has a lower limit at $a>s/2\sim100$~au with a $\gtrsim90\%$ likelihood of having $a<2.5s\sim500$~au. We therefore estimate that the fainter source is likely bound to KELT-24 with an orbital period of order $\sim1000$~yr. This type of signal would only produce a $\sim1$~m~s$^{-1}$~yr$^{-1}$ RV trend via KELT-24, which is too small to measure given the relatively short baseline and large uncertainties in our RVs. Hereafter, we refer to the stellar companion as KELT-24B, and leave the primary as KELT-24. Given the excellent constraints on the pair's relative positions and motions from Gaia, we use the Bayesian rejection sampling routine introduced by the Linear Orbits for the Impatient (LOFTI; \citealt{OFTI, LOFTI}) to estimate the pair's orbit. We allowed our sampling to run until one million trial orbits passed the LOFTI $\chi^2$ probability test, and show the resulting orbital parameters in Figure \ref{fig:KELT24B}. While these accepted trials are not a formal set of posterior orbit solutions, they represent a reasonable starting point for future analyses when additional imaging becomes available, and their convergence reinforces the expectation that the two stars are bound. We estimate a solution consistent with the expectations outlined above, finding a moderately eccentric ($e=0.53$), long-period ($P\approx3000$~yr) orbit with a relatively well-constrained inclination via the instantaneous trajectory derived from Gaia's proper motion measurements. We do, however, exercise caution with this solution, as the proximity and relative brightness between KELT-24 and KELT-24B may be distorting the proper motion measurements from Gaia. Gaia also provides an estimate on the status of KELT-24B as a singular star in the form of their renormalized unit weight error (RUWE). Stars with $\mathrm{RUWE}>1.4$ are generally considered to have poor astrometric solutions or be indicative of binarity, while stars with RUWE of or near unity are good fits to a single-star model. KELT-24B has $\mathrm{RUWE=1.81}$ (KELT-24 has $\mathrm{RUWE=1.014}$), which could be artificially high as a result of its proximity to a much brighter KELT-24, or a clue that the source may be an unresolved binary.


\section{Stellar Parameters} \label{sec:star}

KELT-24 is an early F-type star less than 100~pc from our solar system. To infer the spectral type of KELT-24B, we attempt to derive an uncontaminated color measurement using the resolved NIRC2 and Gaia photometry described in Section \ref{sec:data}. We observe that NIRC2's $K_\mathrm{p}$-band is comparable to the 2MASS $K_\mathrm{s}$-band \citep[e.g.,][]{2010ApJ...718..810S}, allowing us to use the composite magnitude reported by 2MASS and the magnitude difference uncovered by our AO imaging to estimate $G-K_\mathrm{s}$ values of 0.99 and 3.57 for KELT-24 and KELT-24B, respectively. 

\begin{figure}[b]
    \centering
    \includegraphics[width=\linewidth]{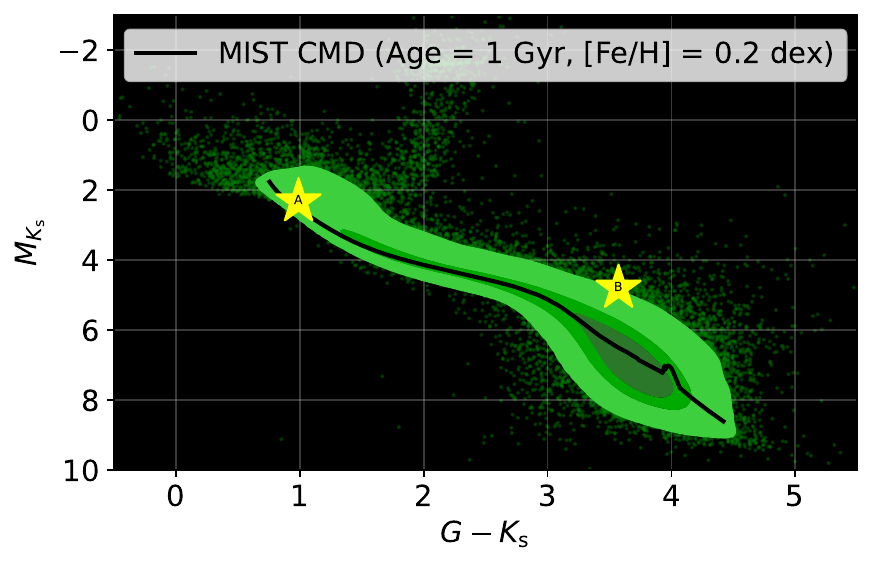}
    \caption{HR diagram of sample of one million stars cross-catalogued between Gaia and 2MASS with density contours. The MIST evolutionary track is shown in black for a 1~Gyr, 0.2~dex metallicity star assuming no extinction. KELT-24 is well-described as an F-type star, while KELT-24B is notably bright for stars of similar color.}
    \label{fig:HR}
\end{figure}

Due to its proximity to KELT-24, however, we must consider the degree to which KELT-24B is blended in the available photometry and spectroscopy used in our analysis. Due the large optical flux ratio between the two stars (see Section \ref{subsec:sed}), the contribution of the companion source to KELT-24's measured RVs are likely to be negligible. The spectrographs used in this analysis described in Section \ref{sec:data}, for example, each have fiber sizes on the order of the angular separation of the pair. As for our photometric observations, we determine that KELT-24B's contribution to the overall flux is $\approx1\%$ in bluer bands like Johnson $B$, and up to $\approx10\%$ in redder bands like 2MASS $K_\mathrm{s}$. Kraus et al. (in prep.) determine that for sources separated by 2'', WISE magnitudes represent full blends, 2MASS magnitudes are only fractionally blended, and high-resolution surveys like Tycho \citep{tycho}, Gaia, and select AO imaging are able to safely resolve the two with little to no photometric interference. However, we determine this effect to be negligible with regard to the deconvolution of the reported $K_\mathrm{s}$ into isolated magnitudes shown above. We depict both targets on a Hertzsprung-Russel (HR) diagram in Figure \ref{fig:HR}. While KELT-24 remains in good agreement with our model for it as an F-type main sequence star, KELT-24B is found to be $\approx3\sigma$ above the median absolute $K_\mathrm{s}$-band magnitude when compared to other stars within a bin of $\pm0.05$ in $G-K_\mathrm{s}$. We therefore estimate that KELT-24B is either an unusually faint M-type star, or more likely an unresolved blend of two stars along the lower main sequence. This possibility is further explored in Section \ref{sec:other_planets}. The assumption of boundness with KELT-24 also suggests that KELT-24B shares its distance and age with the primary. The known identifiers, as well as the observed astrometric and photometric properties of both sources are provided in Table \ref{tab:observed_stellar_params}.


\vspace{1cm}

\startlongtable
\begin{deluxetable*}{lcccc}
\tablecaption{Observed Properties of both stars in the KELT-24 system from the Literature}
\tablehead{\colhead{Parameter} & \colhead{Description} & \colhead{KELT-24} & \colhead{KELT-24B} & \colhead{Reference}}
\startdata
\textbf{Other Identifiers} &  &  &  & \\
MASCARA & ID & 3 & -- &  Hjorth \\
HD & Henry Draper & 93148 & -- &  HD \\
HIP & Hipparcos & 52796 & -- &  Hipparcos \\
TOI & TESS Object of Interest & 1148 & -- & TOI Catalog \\
TIC & TESS Input Catalog & 349827430 & 900697389 & TIC v8 Catalog \\
2MASS & Source ID & J10473836+7139211 & -- & 2MASS \\
Gaia & Source ID & 1076970406751899008 & 1076970406752355584 & Gaia DR3 \\
\textbf{Astrometry} &  & &  & \\
$\alpha_\mathrm{J2016}$ & Right ascension [hh:mm:ss] & 10:47:38.16 & 10:47:38.21 & Gaia DR3 \\
$\delta_\mathrm{J2016}$ & Declination [dd:mm:ss] & +71:39:20.60 & +71:39:18.55 & Gaia DR3 \\
$l$ & Galactic longitude [deg] & 135.5733 & 135.5737 & Gaia DR3 \\
$b$ & Galactic latitude [deg] & 42.3014 & 42.3019 & Gaia DR3 \\
$\pi$ & Parallax [mas] & 10.3218~$\pm$~0.0180 & $10.6799\pm0.0892$ & Gaia DR3 \\
$\mu_{\alpha_\mathrm{J2016}}$ & Proper motion in right ascension [mas~yr$^{-1}$] & -56.061~$\pm$~0.017 &$-49.897\pm0.101$ & Gaia DR3 \\
$\mu_{\delta_\mathrm{J2016}}$ & Proper motion in declination [mas~yr$^{-1}$] & -34.526~$\pm$~0.021& $-37.516\pm0.084$ & Gaia DR3 \\
RV & Absolute radial velocity [km~s$^{-1}$] & -5.749~$\pm$~0.065& -- & Rodriguez \\
$U$ & Space velocity towards galactic center [km~s$^{-1}$] & 	-19.661~$\pm$~0.062 & -16.656~$\pm$~0.220 & Section \ref{subsec:galaxy} \\
$V$ & Space velocity towards galactic spin [km~s$^{-1}$] & 	-22.789~$\pm$~0.056 & -22.466~$\pm$~0.212 & Section \ref{subsec:galaxy} \\
$W$ & Space Velocity towards NGP [km~s$^{-1}$] & 	-6.442~$\pm$~0.044 & -4.332~$\pm$~0.052& Section \ref{subsec:galaxy} \\
\textbf{Magnitudes} &  &  &   & \\
$B$ & Tycho $B$-band magnitude & 	8.82~$\pm$~0.02 & -- & Tycho-2 \\
$V$ & Tycho $V$-band magnitude & 	8.33~$\pm$~0.01 & --  & Tycho-2 \\
$G$ & Gaia $G$-band magnitude & 	8.2486~$\pm$~0.0028 & 13.3030 $\pm$ 0.0030 & Gaia DR3 \\
$J$ & 2MASS $J$-band magnitude & 	7.408~$\pm$~0.024 & --  & 2MASS \\
$H$ & 2MASS $H$-band magnitude & 	7.200~$\pm$~0.044 & --  & 2MASS \\
$K_\mathrm{s}$ & 2MASS $K_\mathrm{s}$-band magnitude & 	7.154~$\pm$~0.021 & -- & 2MASS \\
$K_\mathrm{p}$ & NIRC2 $K_\mathrm{p}$-band magnitude & 	7.260~$\pm$~0.100 & 9.729~$\pm$~0.100 & This work \\
$W1$ & WISE1 magnitude & 7.106~$\pm$~0.032 & -- & WISE \\
$W2$ & WISE2 magnitude & 7.134~$\pm$~0.019 & -- & WISE \\
$W3$ & WISE3 magnitude & 7.148~$\pm$~0.016 & -- & WISE \\
$W4$ & WISE4 magnitude & 7.184~$\pm$~0.082 & -- & WISE \\
\enddata
\tablenotetext{}{Space velocity terms for KELT-24B are calculated assuming the absolute radial velocity observed for KELT-24. References are: Hjorth (H2019), HD \citep{HD}, Hipparcos \citep{Hipparcos}, TOI Catalog \citep{2021ApJS..254...39G}, TIC v8 Catalog \citep{2019AJ....158..138S}, 2MASS \citep{2MASS}, Rodriguez (R2019), Tycho-2 \citep{tycho2}, WISE \citep{2012yCat.2311....0C}}
\label{tab:observed_stellar_params}
\end{deluxetable*}

\vspace{-0.2cm}

\subsection{Spectral Energy Distribution} \label{subsec:sed}

\begin{figure}[h]
    \centering
    \includegraphics[width=\linewidth]{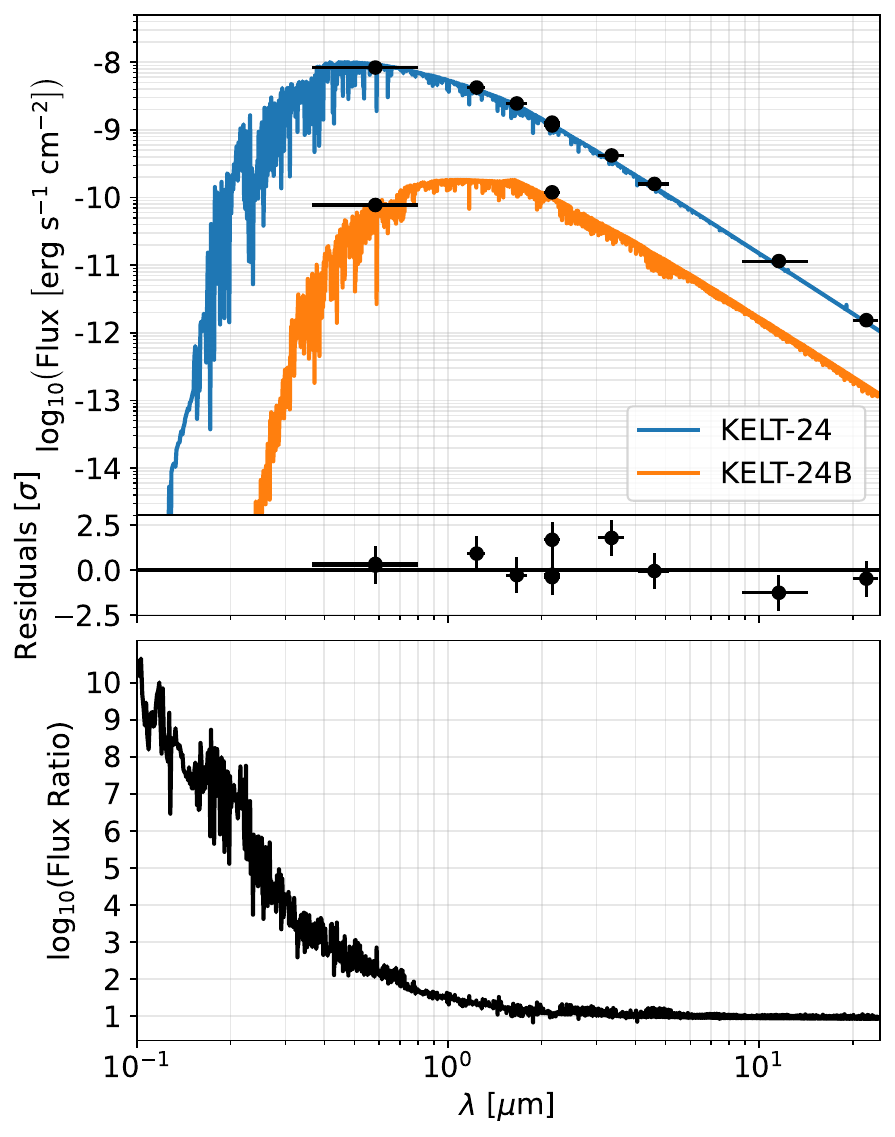}
    \caption{Top: spectral energy distribution of the KELT-24 system modeled for two sources. Black points represent the observed photometric measurements, where the horizontal bars represent the effective width of the bandpass. The blue and orange lines are the corresponding model fluxes from the best-fit atmosphere models for our two sources considered. Bottom: flux ratios of KELT-24 to KELT-24B as a function of wavelength. KELT-24B contributes $<1\%$ of the total flux at bandpasses narrower than $0.7\mu\mathrm{m}$, but is nearly 10\% as bright as KELT-24 in redder bands.} 
    \label{fig:sed}
\end{figure}

For the global analyses described in Section \ref{sec:global_fit}, we use the spectral energy distribution (SED) fitting capabilities within \texttt{EXOFASTv2} \citep{2013PASP..125...83E, 2017ascl.soft10003E, Eastman:2019} to model the stellar properties of KELT-24~based on a known distance, available broadband photometry, and a bolometric flux prior derived via the integration of KELT-24's SED. Similar to the routine executed by R2019, and as an independent determination of the basic stellar parameters, we performed an analysis of the broadband SED of the star together with the Gaia DR3 parallax \citep[with no systematic offset applied; see, e.g.,][]{StassunTorres:2021}, yielding an empirical measurement of the stellar radius following the procedures described in \citet{Stassun:2016,Stassun:2017,Stassun:2018}. We aggregated the $JHK_\mathrm{s}$ magnitudes from 2MASS, the W1--W4 magnitudes from WISE, as well as the resolved Gaia $G$- and NIRC2 Br$\gamma$-bands in order to constrain both the SED of KELT-24 and the contamination of KELT-24B. Together, the available photometry spans the full stellar SED over the wavelength range 0.1--24.1~$\mu$m, and is shown in Figure \ref{fig:sed}. We performed a fit using the MIST bolometric correction tables which were computed from the C3K grid (Conroy et al., in prep) of 1D atmosphere models (based on ATLAS12/SYNTHE; \citealt{1970SAOSR.309.....K, 1979ApJS...40....1K}). We derive the stellar temperature and luminosity from the SED, and use those to infer the stellar radius. The remaining free parameter is the extinction $A_V$, which we limited to the maximum line-of-sight value from the Galactic dust maps of \citet{Schlegel:1998}. Integrating the (unreddened) model SED gives the bolometric flux at Earth, stellar radius, and an estimation of the stellar mass from the empirical relations of \citet{Torres:2010}. In order to constrain the companion flux and its contribution to the SED, we force both it and KELT-24 to have the same ages, initial metallicities, distances, and reddenenings.



\vspace{-0.2cm}
\subsection{Age \& Galactic Kinematics} \label{subsec:galaxy}

From the spectroscopic $v\sin i$ and stellar radius, we estimate the stellar rotation period of KELT-24 to be $P_{\rm rot} / \sin i \sim4$~d. While, for example, following the gyrochronology relations from \citet{Barnes:2007} yields a relatively young ($\approx1~\mathrm{Gyr}$) age estimate for KELT-24 that is consistent with the results of both R2019 and H2019, \cite{2020ApJ...904..140C} show that KELT-24 is in a regime of high effective temperatures that does not lend itself to this type of age determination. Therefore, we employ MIST evolutionary models to estimate the stellar age in our joint analyses described in Section \ref{sec:global_fit}. Furthermore, we fix the age of both sources considered in our fits to be equal.



The enhancement in KELT-24's astrometric precision since the introduction of Gaia eDR3, which was released after KELT-24~b's discovery, allow us to place more rigid constraints on KELT-24's space motion and place in the galaxy. Following the public BANYAN tool \citep{2018ApJ...856...23G}, we estimate KELT-24 to have a 99.9\% probability of being a field star, and not a member of any larger cluster. According to Appendix A of \cite{2014A&A...562A..71B}, KELT-24 has a $>99\%$ likelihood of being in the thin disk.

We used \texttt{galpy} \citep{2015ApJS..216...29B} to determine the three-dimensional absolute \textit{UVW} galactic space motion of KELT-24, given in Table \ref{tab:observed_stellar_params}. Adopting the results from \cite{2011MNRAS.412.1237C} for thin-disk dwarfs, we determine KELT-24's galactic space motion with respect to the local standard of rest to be $U=-11.161\pm0.297$~km~s$^{-1}$, $V=-9.409\pm0.434$~km~s$^{-1}$, and $W=0.048\pm0.264$~km~s$^{-1}$. KELT-24's position in the galaxy, combined with its relatively low absolute galactic space motion, substantiates its inferred young age.

\section{Global Fitting} \label{sec:global_fit}

In order to retrieve the KELT-24 system parameters, we jointly model the observed transits and measured RVs with \texttt{EXOFASTv2} \citep{2013PASP..125...83E, 2017ascl.soft10003E, Eastman:2019}. However, we found the derived stellar parameters to be highly-discrepant, at the level of $>5\sigma$, with KELT-24's physical properties previously reported in the literature, specifically R2019 and H2019.


To address these discrepancies, we attempted to recover the stellar parameters of KELT-24 via three different approaches. The first two derivations of KELT-24's physical properties were designed to mirror the processes of R2019 and H2019 using our updated datasets. We then directed a fit to determine the stellar parameters by excluding the SED. In general, stellar parameters for a transiting planet can be inferred by imposing an agreement between evolutionary models, the SED, and transit-derived properties. Each method outlined here should return a set of stellar parameters consistent with the others. The fact that this is not the case implies that one or more of our model assumptions is incorrect. 


A major factor in the determination of a star's physical properties is the reliability of the inferred stellar density. The majority of our transit light curves come from high-precision TESS photometry, meaning the constraint on the stellar density from the transits should be trustworthy. However, there are several known processes that could bias the stellar density. For example, unaccounted flux contribution to a transit light curve from an unseen companion will cause an underestimation of the stellar density. Following Equation 9 from \cite{2014MNRAS.440.2164K} using our transit-derived values, we find that an unresolved source contributing an additional 90\% of relative flux would be needed to bridge the discrepancies in the stellar density seen here. While an unresolved source of this magnitude can be ruled out, KELT-24B is known to contribute $\approx1-10\%$ of KELT-24's flux depending on the bandpass, so we further consider the possibility of fainter sources that could at least partially be blending our transit light curves by executing a method that excludes the SED in Section \ref{subsec:no_sed}. \cite{2014MNRAS.440.2164K} also describes the effect that transit timing variations (TTVs) may have on the stellar density estimation. According to Equation 23 of their work, a TTV amplitude of $A_\mathrm{TTV}\approx3.2~\mathrm{minutes}$ would be enough to induce the apparent underestimation of the stellar density that we observe (see Table \ref{tab:derived_stellar_params}). While TTVs could be present as a result of some distant companion orbiting KELT-24, we do not detect any signals of TTVs at or near this level (see Section \ref{sec:other_planets}). It is also possible that a bias in eccentricity could bias the transit duration and impact the inferred stellar density. For example, following Equations 33 and 34 from \cite{2014MNRAS.440.2164K}, even the discrepancy in eccentricity between R2019 ($e=0.077^{+0.024}_{-0.025}$) and H2019 ($e=0.050^{+0.020}_{-0.017}$) is enough to bias the stellar density by up to 10\%. Each of our trial analyses find a lower best-fit value for the orbital eccentricity of KELT-24~b, an indicator that the eccentricity may still be biased high \citep[e.g.,][]{2006ApJ...646..505B, 2011MNRAS.410.1895Z}. In fact, even statistical analyses of planets with circular ($e=0$) orbits may return a nonzero eccentricity as a result of the Lucy-Sweeney bias \citep{1971AJ.....76..544L}, meaning that KELT-24~b could have circularized but require additional observations to reconcile that. We also attempted to exploit the flicker method \citep{2014ApJ...785L..32K, 2016ApJ...818...43B} as an independent effort to determine the stellar density, but were unable to apply it to KELT-24's light curve from TESS.

The default evolutionary track modeling used within \texttt{EXOFASTv2} is MIST \citep{MIST0, MIST1}, which has been shown to be dependable for isolated, main sequence stars like KELT-24, but does not reliably reproduce low-mass stars. Systematic errors in the evolutionary model consistent with those identified by \cite{2022ApJ...927...31T} are assumed, but it is possible that KELT-24 could violate our theoretical expectation for such a star. This leaves the SED as the remaining possible source for confusion. The only known nearby source to KELT-24 is KELT-24B, for which we have two resolved photometric observations described in Section \ref{subsec:sed}. All other photometry used in the SED are composite magnitudes of both sources, and it is possible that there is an unresolved stellar multiple masked as a single star that we are unable to account for. As for TESS, crowding in the photometric aperture is corrected for each target star and sector in the SPOC Presearch Data Conditioning pipeline module based on a “crowding metric” computed for the photometric aperture pixels selected in the Compute Optimal Aperture pipeline module (see \citealt{coa1, coa2} regarding selection of photometric aperture pixels and computation of the crowding metric). The crowding metric represents the time-averaged (by sector) fraction of flux in the photometric aperture attributable to the target star itself (after background subtraction). The crowding metric computation in each sector employs the CCD-specific Pixel Response Function to model the point spread function and pixel response at the location of the target and nearby stars that can contribute flux to the photometric aperture (see \citealt{pdc_cor1, pdc_cor2} for a complete description of the pipeline crowding correction). In the case of KELT-24, 1.4\% of the median flux level over all cadences in each sector is then subtracted from the light curve, and reported in the PDCSAP fluxes, to account for contaminating sources, namely KELT-24 B. We probe the KELT-24 system for additional, unseen companions in Section \ref{sec:other_planets}, but do not find evidence for anything bright enough to contaminate the SED in a meaningful way. Though none of the fits presented here in this section can definitively be deemed the most correct, we propose that the SED is the most likely culprit for the discrepancies in stellar parameters previously reported. For the purpose of publishing a best-fit set of system parameters, we promote a global analysis for the KELT-24 system that neglects the SED.

We begin by modeling the KELT-24 system following the individual routines executed by R2019 (Section \ref{subsec:rodriguez_reproduce}) and H2019 (Section \ref{subsec:mascara_reproduce}) with the goal of reproducing their respective sets of system posteriors, and by extension their apparent physical disparities. The method published by R2019 was a joint analysis of both KELT-24 and KELT-24~b using all available data, incorporating both the SED and associated blend corrections to account for KELT-24B, here assumed to be a single source. H2019, however, modeled the star from the SED in isolation and subsequently derived the physical properties of KELT-24~b using transit photometry, RVs, and a fixed stellar template. In principle, both of these methods should have recovered the same stellar parameters for KELT-24. As discussed below, our reproductions of R2019 and H2019 are reasonably consistent with the original discovery papers, yet again inconsistent with each other, further suggesting that there is an underlying systematic issue in play. The results of Section \ref{subsec:rodriguez_reproduce} (e.g., R2019) could be skewed if we are misinterpreting the transit light curves, perhaps via an unknowingly blended source, for example. Meanwhile, the results of Section \ref{subsec:mascara_reproduce} (e.g., H2019) could be erroneous if the SED is in some way contaminated. Presumably, there is an inherent disconnect between KELT-24's transit-derived stellar parameters and those modeled based on the SED. In an attempt to isolate this complication, we finally employ an additional analysis that jointly models the KELT-24 system without the SED, leaving the MIST evolutionary model and observed data as the drivers of KELT-24's system parameters.

The priors on KELT-24 used for each fit are given in Table \ref{tab:priors_all}. Fits that do not include the SED are supplied a spectroscopic effective temperature prior from our TRES observations, since that is otherwise derived from an SED. The prior placed on KELT-24~b's time of inferior conjunction is derived from KELT photometry, and serves to extend the observed baseline and more precisely constrain the planetary period. We exclude priors on Claret limb-darkening coefficients \citep{claret}, allowing them to float freely. This allows for a more unbiased determination of the stellar temperature. For each of the three analyses we have carried out, we jointly model the parameters for both KELT-24 and KELT-24B, but fix both stars' distances, ages, reddenings, initial metallicities, and SED error scalings to be equal to each other, using a public but experimental branch of \texttt{EXOFASTv2}. These are reasonable allowances given the high-likelihood of binarity of the two stars. For each global fit, we include all available photometric and spectroscopic observations discussed in Section \ref{sec:data}. Each method simultaneously fits for the DT signals observed by MINERVA and TRES on 2019 March 31. Detailed descriptions of each of these three analyses, as well as their resulting estimates for KELT-24's stellar parameters, are subsequently described below, and summarized in Table \ref{tab:derived_stellar_params}. The derived properties of KELT-24~b are provided for each analysis in Table \ref{tab:planet_posteriors}. 
\vspace{-0.35cm}
\startlongtable
\begin{deluxetable}{lccc}
\tablecaption{Priors in the \texttt{EXOFASTv2} Global Fits for KELT-24}
\tablehead{\colhead{Param.} & \colhead{Description} & \colhead{Prior}}
\startdata
$\varpi$ & Parallax [mas] & $\mathcal{N}$(10.3620, 0.04165) \\
$T^*_\mathrm{eff}$ & Eff. Temp. [K] & $\mathcal{N}$(6499, 98) \\
$\left[\mathrm{Fe/H}\right]$ & Metallicity [dex] & $\mathcal{N}$(0.16, 0.08) \\ 
$v\sin i$ & Rot. Vel. [$\mathrm{km~s^{-1}}$] & $\mathcal{N}$(19.46, 0.18) \\
$A_V$ & Reddening [mag] & $\mathcal{U}$[0, 0.16337] \\
$T_\mathrm{c}$ & Inf. Conj. [d] & $\mathcal{N}$(2457147.0522, 0.0021) \\
\enddata
\tablenotetext{}{*Spectroscopically-derived, and only used for the case where the SED is excluded. Here, $\mathcal{N}$(0,1) is a Gaussian distribution centered at zero with a standard deviation of one, and $\mathcal{U}$[0, 1] is a uniform distribution bounded inclusively between zero and one.}
\label{tab:priors_all}
\end{deluxetable}

\subsection{Global Modeling Using SED} \label{subsec:rodriguez_reproduce} 

First, we performed a global fit to parameterize the KELT-24 system, including KELT-24B identified in Section \ref{sec:imaging}. To do this, we employ a new functionality of \texttt{EXOFASTv2} to model the observed SED as a superposition of the two stars captured in the broadband resolved photometry. We therefore do not enforce a prior on KELT-24's effective stellar temperature, as it will be derived from the SED. The analysis presented here differs from that of R2019 in that we do not enforce priors derived from an independent analysis of the SED on $T_\mathrm{eff}$ and $R_*$. Instead, we jointly fit for KELT-24's stellar parameters using the SED, MIST models, and KELT-24~b's transits. A stark discrepancy is observed between the $T_\mathrm{eff}$ and $R_*$ inferred from this combination when compared to those derived directly from the SED. The discrepancies between $T_\mathrm{eff}$/$T_\mathrm{eff_{SED}}$ and $R_\mathrm{*}$/$R_\mathrm{*_{SED}}$ (see Table \ref{tab:derived_stellar_params}) highlight a troubling disagreement between the three independent models used to inform our stellar parameters. This subtle inconsistency was less apparent in R2019 because of the aforementioned independently derived priors on $T_\mathrm{eff}$ and $R_*$ from the SED. Here, we find a stellar density for KELT-24 of $\rho_*=0.709\pm0.031\mathrm{~g~cm^{-3}}$.

\subsection{Piecewise Modeling Using SED} \label{subsec:mascara_reproduce}

As a second effort to recover the parameters of the KELT-24 system, we carry out a subsequent analysis outlined by H2019 in their near-simultaneous discovery of KELT-24~b as MASCARA-3~b. To do this, we exclude transits and therefore do not allow the light curve to inform the stellar fit. Rather, we independently model the star using the blended SED with \texttt{EXOFASTv2} just as in Section \ref{subsec:rodriguez_reproduce}, as well as MIST. The SED-derived parameters, while consistent with those from Section \ref{subsec:rodriguez_reproduce}, are again discrepant with the overall model's best-fit stellar properties (see Table \ref{tab:derived_stellar_params}). We then model the transit light curve with wide, uniform priors on the stellar parameters. We extract just the transit observables and combine them with the stellar observables to derive the complete set of system parameters. This method returns a stellar density of $\rho_*=0.501^{+0.077}_{-0.081}~\mathrm{g~cm^{-3}}$, which differs significantly from the value determined from transits alone ($\rho_{*_\mathrm{transit}}=0.746\pm0.033~\mathrm{g~cm^{-3}}$). Given the number of high-precision, background-subtracted TESS transits available for KELT-24~b, and since none of the effects described earlier in this section are likely to bias the stellar density at the level of discrepancy seen here, the transit-derived stellar density should be trustworthy, As such, this disagreement between $\rho_*$ and $\rho_{*_\mathrm{transit}}$ is a concern. This discrepancy between the transit-derived stellar density and jointly modeled stellar density was also present in H2019, but was not explored.
\vspace{1cm}
\subsection{No SED Contribution} \label{subsec:no_sed}

\begin{figure*}
    \centering
    \includegraphics[width=\linewidth]{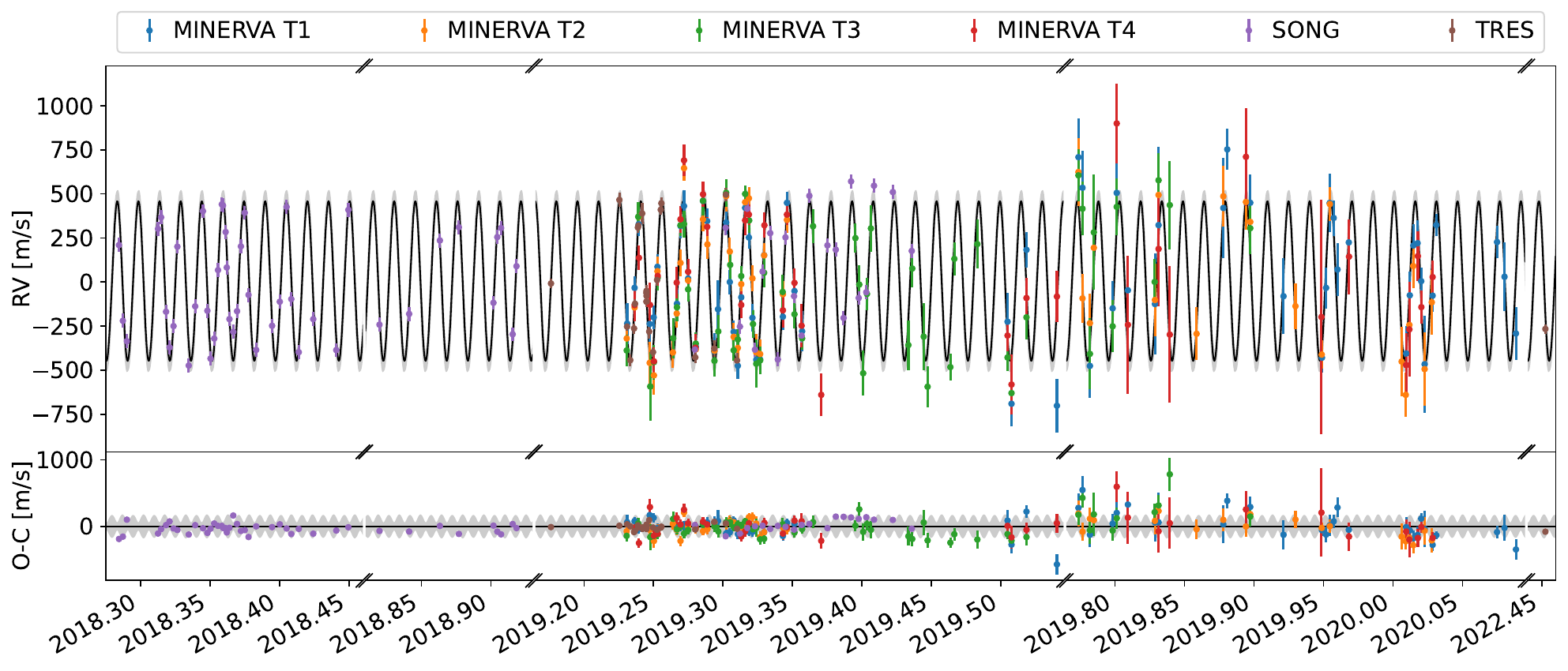}
    \caption{Full RV timeseries including all 291 RVs used in our analysis.}
    \label{fig:allRV_timeseries}
\end{figure*}

\begin{figure}
    \centering
    \includegraphics[width=\linewidth]{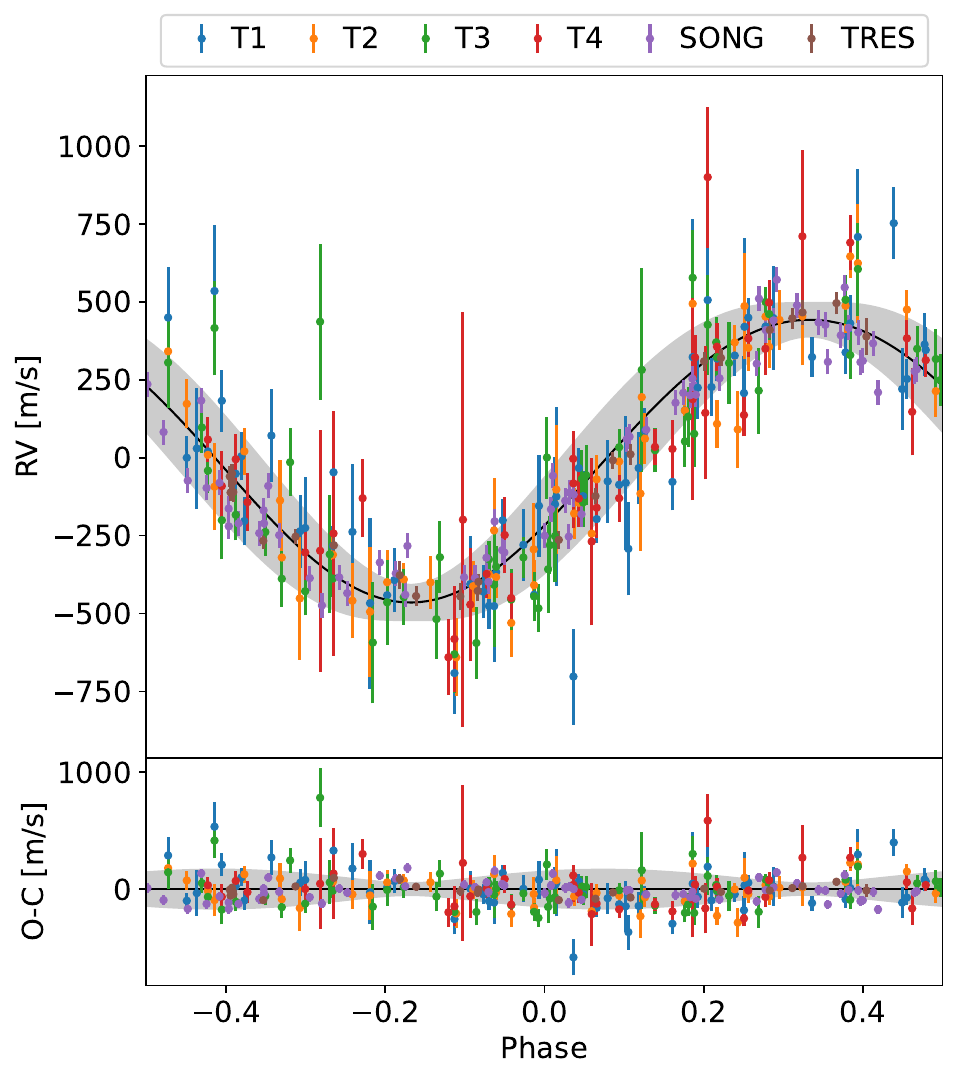}
    \caption{Phase-folded diagram of all 291 RVs used in our analysis. T1-T4 are the four MINERVA telescopes.}
    \label{fig:allRV_phase}
\end{figure}

\begin{figure}
    \centering
    \includegraphics[width=\linewidth]{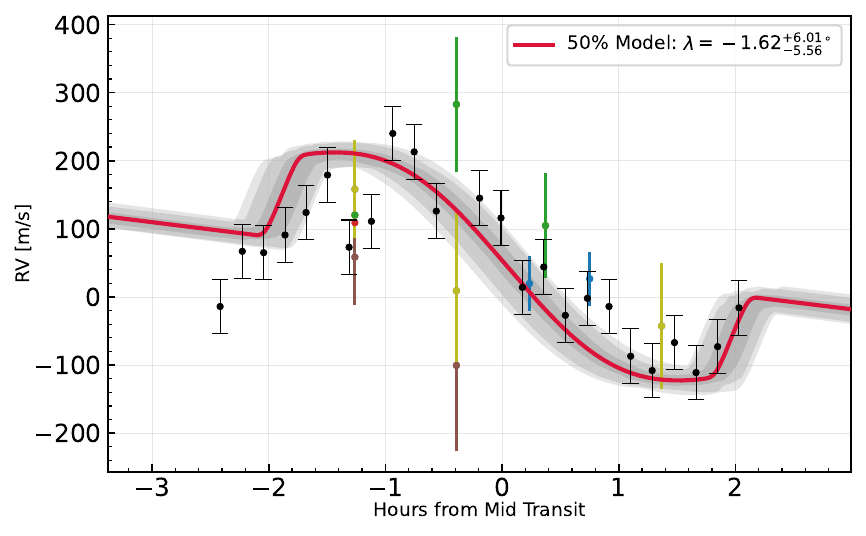}
    \caption{Independent determination of the RM signal for KELT-24~b using only the in-transit SONG spectra taken with the ThAr cell. The best-fit spin-orbit angle (${\lambda=-1.62^{+6.01}_{-5.56}}^{\circ}$) is consistent with our findings presented in Table \ref{tab:planet_posteriors}. The gray regions represent the 1-, 2-, and 3-$\sigma$ deviations from our best-fit model. The spurious in-transit RVs from our analysis are simply overlaid to demonstrate agreement, and are not included in the fit here.}
    \label{fig:RVRM}
\end{figure}

Here, we consider the possibility that the SED for KELT-24 is corrupted, perhaps as a result of stellar blending from unresolved sources. Since this method excludes the SED, we enforce a spectroscopic prior on $T_\mathrm{eff}$, allowing the MIST models and transit photometry to inform our constraints on the remaining stellar parameters. Much like Section \ref{subsec:rodriguez_reproduce}, the stellar and planetary parameters are jointly modeled. We appoint the SED as the most likely source for the discrepancies observed, and hereafter adopt this model as our favored procedure for globally fitting the KELT-24 system to derive its system parameters. The best-fit orbit is shown through a full RV timeseries and phase-folded RV curve in Figure \ref{fig:allRV_timeseries} and Figure \ref{fig:allRV_phase}, respectively. While they are not included in our global models presented here, we independently fit for the RM signal of KELT-24~b using the in-transit RVs collected from SONG using the ThAr cell. The best-fit model from \texttt{rmfit} \citep{2022ApJ...931L..15S} is shown in Figure \ref{fig:RVRM}.

\vspace{-0.5cm}
\startlongtable
\begin{deluxetable*}{llcc>{\columncolor{celadon}}c} \label{tab:derived_stellar_params}
\tablecaption{Median values and 68\% confidence interval for KELT-24's stellar parameters}
\tablehead{\colhead{~~Parameter} & \colhead{Description} & \colhead{Global Modeling (Sec. \ref{subsec:rodriguez_reproduce})}  & \colhead{Piecewise Modeling (Sec. \ref{subsec:mascara_reproduce})}  & \colhead{No SED (Sec. \ref{subsec:no_sed})}}
\startdata
~~~~$M_*$\dotfill &Mass [\msun] \dotfill &$1.372^{+0.049}_{-0.050}$&$1.293^{+0.072}_{-0.11}$&$1.268\pm0.063$\\
~~~~$R_*$\dotfill &Radius [\rsun]\dotfill &$1.397^{+0.027}_{-0.026}$&$1.537^{+0.068}_{-0.061}$&$1.338^{+0.032}_{-0.030}$\\
~~~~$R_\mathrm{*,~SED}$\dotfill &Radius$^1$ [\rsun]\dotfill &$1.474\pm0.019$&$1.493^{+0.023}_{-0.021}$&$--$\\
~~~~$L_*$\dotfill &Luminosity [\lsun]\dotfill &$3.41^{+0.19}_{-0.16}$&$3.47^{+0.20}_{-0.16}$&$2.75^{+0.24}_{-0.21}$\\
~~~~$F_\mathrm{Bol}$\dotfill &Bolometric Flux [cgs]\dotfill &$0.00000001174^{+0.00000000065}_{-0.00000000055}$&$0.00000001191^{+0.00000000068}_{-0.00000000054}$&--\\
~~~~$\rho_*$\dotfill &Density [cgs]\dotfill &$0.709\pm0.031$&$0.501^{+0.077}_{-0.081}$&$0.744^{+0.032}_{-0.033}$\\
~~~~$\log{g}$\dotfill &Surface Gravity [cgs]\dotfill &$4.285^{+0.013}_{-0.014}$&$4.176^{+0.046}_{-0.061}$&$4.287\pm0.014$\\
~~~~$T_{\rm eff}$\dotfill &Effective Temperature [K]\dotfill &$6639^{+94}_{-91}$&$6360\pm150$&$6426^{+93}_{-91}$\\
~~~~$T_{\rm eff,~SED}$\dotfill &Effective Temperature$^1$ [K]\dotfill &$6493^{+100}_{-89}$&$6443^{+110}_{-88}$&--\\
~~~~$[{\rm Fe/H}]$\dotfill &Metallicity [dex]\dotfill &$0.160\pm0.078$&$0.126\pm0.081$&$0.138^{+0.086}_{-0.082}$\\
~~~~$[{\rm Fe/H}]_{0}$\dotfill &Initial Metallicity$^{2}$ [dex]\dotfill &$0.230^{+0.064}_{-0.058}$&$0.204^{0.073}_{-0.077}$&$0.193^{+0.066}_{-0.067}$\\
~~~~Age\dotfill &Age [Gyr]\dotfill &$1.05^{+0.54}_{-0.40}$&$3.0^{+2.2}_{-1.2}$&$1.98^{+0.95}_{-0.79}$\\
~~~~EEP\dotfill &Equal Evolutionary Phase$^{3}$ \dotfill &$328^{+11}_{-13}$&$376^{+46}_{-30}$&$345^{+20}_{-15}$\\
~~~~$v\sin I_*$\dotfill &Projected Rotational Velocity [km~s$^{-1}$]\dotfill &$19.75^{+0.13}_{-0.15}$&$19.75^{+0.13}_{-0.14}$&$19.74^{+0.13}_{-0.15}$\\
~~~~$V_\mathrm{line}$\dotfill &Unbroadened Line Width [km~s$^{-1}$]\dotfill &$5.77^{+0.50}_{-0.48}$&$5.78^{+0.50}_{-0.48}$&$5.80^{+0.50}_{-0.53}$\\
~~~~$A_V$\dotfill &$V$-band extinction [mag] \dotfill &$0.068^{+0.058}_{-0.047}$&$0.057^{+0.062}_{-0.042}$&--\\
~~~~$\sigma_{\rm SED}$\dotfill & SED photometry error scaling\dotfill &$1.17^{+0.43}_{-0.27}$&$1.20^{+0.52}_{-0.29}$&--\\
~~~~$\dot{\gamma}$\dotfill &RV slope$^{4}$ [m~s$^{-1}$~d$^{-1}$]\dotfill &$0.025^{+0.029}_{-0.025}$&$0.019^{+0.029}_{-0.025}$&$0.019^{+0.029}_{-0.024}$\\
\enddata
\tablenotetext{}{Stellar parameters for both KELT-24 and KELT-24B, respectively denoted by indices A and B, derived from the three efforts outlined above. The column of stellar parameters we favor from our analyses is highlighted in green.}
\tablenotetext{1}{This value ignores the systematic error and is for reference only}
\tablenotetext{2}{The metallicity of the star at birth}
\tablenotetext{3}{Corresponds to static points in a star's evolutionary history. See \S2 in \citet{MIST0}.}
\tablenotetext{4}{Reference epoch = 2458984.524152}
\end{deluxetable*}
~
~
\startlongtable
\begin{deluxetable*}{llcc>{\columncolor{celadon}}c} \label{tab:planet_posteriors}
\tabletypesize{\small}
\tablecaption{Median values and 68\% confidence interval for KELT-24~b}
\tablehead{\colhead{~~~Parameter} & \colhead{Description} & \colhead{Section \ref{subsec:rodriguez_reproduce}}  & \colhead{Section \ref{subsec:mascara_reproduce}}  & \colhead{Section \ref{subsec:no_sed}}}
\startdata
~~~~$P$\dotfill &Period [d]\dotfill &$5.55149297^{+0.00000052}_{-0.00000053}$&\periodref&$5.55149296^{+0.00000050}_{-0.00000053}$\\
~~~~$R_P$\dotfill &Radius [\rj]\dotfill &$1.183^{+0.023}_{-0.022}$&\rpone&$1.134^{+0.027}_{-0.026}$\\
~~~~$M_P$\dotfill &Mass [\mj]\dotfill &$4.85\pm0.14$&\mpone&$4.59\pm0.18$\\
~~~~$T_C$\dotfill &Time of conjunction$^{1}$ [\bjdtdb]\dotfill &$7147.05249\pm0.00020$&\tcref&$7147.05250\pm0.00019$\\
~~~~$T_T$\dotfill &Min. projected separation$^{2}$ [\bjdtdb]\dotfill &$7147.05249\pm0.00020$&\ttref&$7147.05250\pm0.00019$\\
~~~~$T_0$\dotfill &Optimal conj. Time$^{3}$ [\bjdtdb]\dotfill &$9184.450410\pm0.000035$&\tzeroref&$9173.347426\pm0.000034$\\
~~~~$a$\dotfill &Semi-major axis (au)\dotfill &$0.06826^{+0.00080}_{-0.00084}$&\aone&$0.0665\pm0.0011$\\
~~~~$i$\dotfill &Inclination [deg]\dotfill &$89.56^{+0.31}_{-0.34}$&\iref&$89.67^{+0.24}_{-0.36}$\\
~~~~$e$\dotfill &Eccentricity \dotfill &$0.0356^{+0.011}_{-0.0087}$&$0.0319^{+0.0080}_{-0.0074}$&$0.0319^{+0.0079}_{-0.0074}$\\
~~~~$\omega_*$\dotfill &Argument of Periastron [deg]\dotfill &$29^{+18}_{-24}$&$0^{+28}_{-29}$&$3^{+26}_{-27}$\\
~~~~$T_{\rm eq}$\dotfill &Equilibrium temperature$^{4}$ [K]\dotfill &$1448^{+19}_{-17}$&\teqone&$1391^{+22}_{-23}$\\
~~~~$\tau_{\rm circ}$\dotfill &Tidal circularization timescale [Gyr]\dotfill &$13.1\pm1.1$&\tcircone&$14.6^{+1.1}_{-1.2}$\\
~~~~$K$\dotfill &RV semi-amplitude [$\mathrm{m~s^{-1}}$]\dotfill &$450.4^{+7.5}_{-7.7}$&\kref&$449.1^{+7.2}_{-7.7}$\\
~~~~$R_P/R_*$\dotfill &Radius of planet in stellar radii \dotfill &$0.087019^{+0.00013}_{-0.000099}$&\rprsref&$0.087024^{+0.00012}_{-0.000093}$\\
~~~~$a/R_*$\dotfill &Semi-major axis in stellar radii \dotfill &$10.51^{+0.15}_{-0.16}$&\arref&$10.67^{+0.15}_{-0.16}$\\
~~~~$\delta$\dotfill &$\left(R_P/R_*\right)^2$ \dotfill &$0.007572^{+0.000022}_{-0.000017}$&\deltaref&$0.007573^{+0.000021}_{-0.000016}$\\
~~~~$\delta_{\rm R}$\dotfill &Transit depth in R [fraction]\dotfill &$0.00883^{+0.00024}_{-0.00023}$&\deltarref&$0.00890^{+0.00024}_{-0.00022}$\\
~~~~$\delta_{\rm i'}$\dotfill &Transit depth in i' [fraction]\dotfill &$0.00880^{+0.00021}_{-0.00020}$&$0.00880^{+0.00021}_{-0.00020}$&$0.00883^{+0.00022}_{-0.00021}$\\
~~~~$\delta_{\rm z'}$\dotfill &Transit depth in z' [fraction]\dotfill &$0.00817^{+0.00017}_{-0.00016}$&\deltazref&$0.00831^{+0.00018}_{-0.00017}$\\
~~~~$\delta_{\rm TESS}$\dotfill & TESS fractional transit depth \dotfill &$0.008727^{+0.000087}_{-0.000086}$&\deltatref&$0.008729^{+0.000087}_{-0.000081}$\\
~~~~$\tau$\dotfill &Ingress/egress transit dur. [d]\dotfill &$0.014474^{+0.00021}_{-0.000086}$&\tauref&$0.014437^{+0.00019}_{-0.000053}$\\
~~~~$T_{14}$\dotfill &Total transit duration [d]\dotfill &$0.17934^{+0.00017}_{-0.00016}$&\tonefourref&$0.17930\pm0.00016$\\
~~~~$T_{FWHM}$\dotfill &FWHM transit dur. [d]\dotfill &$0.16482\pm0.00015$&\tfwhmref&$0.16481^{+0.00014}_{-0.00015}$\\
~~~~$b$\dotfill &Transit impact parameter \dotfill &$0.078^{+0.062}_{-0.055}$&\bref&$0.061^{+0.066}_{-0.045}$\\
~~~~$b_S$\dotfill &Eclipse impact parameter \dotfill &$0.081^{+0.063}_{-0.057}$&$0.059^{+0.059}_{-0.042}$&$0.061^{+0.066}_{-0.045}$\\
~~~~$\tau_S$\dotfill &Ingress/egress eclipse dur. [d]\dotfill &$0.01502^{+0.00047}_{-0.00043}$&$0.01447^{+0.00047}_{-0.00045}$&$0.01455^{+0.00046}_{-0.00041}$\\
~~~~$T_{S,14}$\dotfill &Total eclipse duration [d]\dotfill &$0.1854^{+0.0055}_{-0.0050}$&$0.1793^{+0.0055}_{-0.0053}$&$0.1799^{+0.0053}_{-0.0048}$\\
~~~~$T_{S,FWHM}$\dotfill &FWHM eclipse duration [d]\dotfill &$0.1703^{+0.0051}_{-0.0046}$&$0.1648^{+0.0050}_{-0.0049}$&$0.1653^{+0.0049}_{-0.0044}$\\
~~~~$\delta_{S,2.5\mu m}$\dotfill & 2.5$\mu$m blackbody eclipse dep. [ppm]\dotfill &$200.3^{+7.2}_{-6.7}$&$180.5^{+9.7}_{-9.4}$&$178.2^{+8.8}_{-8.9}$\\
~~~~$\delta_{S,5.0\mu m}$\dotfill & 5.0$\mu$m blackbody eclipse dep. [ppm]\dotfill &$652\pm11$&$621\pm16$&$618\pm15$\\
~~~~$\delta_{S,7.5\mu m}$\dotfill & 7.5$\mu$m blackbody eclipse dep. [ppm]\dotfill &$918^{+12}_{-11}$&$887\pm16$&$886\pm15$\\
~~~~$\rho_P$\dotfill &Density [cgs]\dotfill &$3.63\pm0.17$&\rhopone&$3.90^{+0.19}_{-0.20}$\\
~~~~$logg_P$\dotfill &Surface gravity [cgs]\dotfill &$3.934^{+0.014}_{-0.015}$&\loggpone&$3.946\pm0.014$\\
~~~~$\lambda$\dotfill &Projected obliquity [deg]\dotfill &$1.6^{+4.5}_{-3.9}$&$1.7^{+5.2}_{-5.0}$&$1.4^{+4.9}_{-4.7}$\\
~~~~$\Theta$\dotfill &Safronov Number \dotfill &$0.4075^{+0.0100}_{-0.0099}$&$0.421^{+0.025}_{-0.021}$&$0.424^{+0.011}_{-0.012}$\\
~~~~$\fave$\dotfill &Incident Flux [\fluxcgs]\dotfill &$0.997^{+0.052}_{-0.047}$&\faveone&$0.849^{+0.055}_{-0.054}$\\
~~~~$T_P$\dotfill &Time of Periastron [\bjdtdb]\dotfill &$2457146.17^{+0.27}_{-0.37}$&$2457145.71^{+0.43}_{-0.46}$&$2457145.77^{+0.39}_{-0.43}$\\
~~~~$T_S$\dotfill &Time of eclipse [\bjdtdb]\dotfill &$2457144.381\pm0.025$&\tsref&$2457144.379^{+0.024}_{-0.023}$\\
~~~~$T_A$\dotfill &Time of Ascending Node [\bjdtdb]\dotfill &$2457151.297^{+0.030}_{-0.028}$&$2457151.266^{+0.030}_{-0.029}$&$2457151.270^{+0.030}_{-0.029}$\\
~~~~$T_D$\dotfill &Time of Descending Node [\bjdtdb]\dotfill &$2457148.461^{+0.027}_{-0.028}$&$2457148.491^{+0.030}_{-0.029}$&$2457148.488\pm0.027$\\
~~~~$V_c/V_e$\dotfill & \dotfill &$0.983\pm0.014$&$1.000\pm0.015$&$0.998\pm0.014$\\
~~~~$e\cos{\omega_*}$\dotfill & \dotfill &$0.0296^{+0.0071}_{-0.0070}$&$0.0285^{+0.0070}_{-0.0069}$&$0.0288^{+0.0069}_{-0.0064}$\\
~~~~$e\sin{\omega_*}$\dotfill & \dotfill &$0.017^{+0.015}_{-0.014}$&$0.000\pm0.015$&$0.002^{+0.015}_{-0.014}$\\
~~~~$M_P\sin i$\dotfill &Minimum mass [\mj]\dotfill &$4.85\pm0.14$&\msinione&$4.59\pm0.18$\\
~~~~$M_P/M_*$\dotfill &Mass ratio \dotfill &$0.003376^{+0.000072}_{-0.000070}$&$0.00342^{+0.00020}_{-0.00016}$&$0.003457\pm0.000078$\\
~~~~$d/R_*$\dotfill &Separation at mid transit \dotfill &$10.32^{+0.29}_{-0.30}$&$10.69^{+0.33}_{-0.32}$&$10.65^{+0.30}_{-0.31}$\\
~~~~$P_T$\dotfill &A priori non-grazing transit prob \dotfill &$0.0885^{+0.0027}_{-0.0024}$&$0.0854^{+0.0027}_{-0.0026}$&$0.0858^{+0.0026}_{-0.0023}$\\
~~~~$P_{T,G}$\dotfill &A priori transit prob \dotfill &$0.1053^{+0.0032}_{-0.0029}$&$0.1017^{+0.0032}_{-0.0030}$&$0.1021^{+0.0031}_{-0.0028}$\\
~~~~$P_S$\dotfill &A priori non-grazing eclipse prob \dotfill &$0.08545^{+0.00052}_{-0.00023}$&$0.08535^{+0.00039}_{-0.00017}$&$0.08534^{+0.00047}_{-0.00018}$\\
~~~~$P_{S,G}$\dotfill &A priori eclipse prob \dotfill &$0.10174^{+0.00064}_{-0.00028}$&$0.10161^{+0.00049}_{-0.00020}$&$0.10160^{+0.00059}_{-0.00021}$\\
\multicolumn{2}{l}{Wavelength Parameters:}&&&\smallskip \\
~~~~$u_{1,~R}$\dotfill &$R$-band linear limb-darkening\dotfill &$0.288^{+0.045}_{-0.046}$&\ruoneref&$0.300^{+0.045}_{-0.043}$\\
~~~~$u_{2,~R}$\dotfill &$R$-band quadratic limb-darkening \dotfill &$0.294\pm0.048$&\rutworef&$0.298\pm0.048$\\
~~~~$u_{1,~i'}$\dotfill &$i$-band linear limb-darkening\dotfill &$0.282\pm0.040$&$0.282^{+0.040}_{-0.041}$&$0.288^{+0.043}_{-0.041}$\\
~~~~$u_{2,~i'}$\dotfill &$i$-band quadratic limb-darkening\dotfill &$0.323\pm0.046$&$0.322\pm0.046$&$0.336^{+0.046}_{-0.047}$\\
~~~~$u_{1,~z'}$\dotfill &$z$-band linear limb-darkening \dotfill &$0.147^{+0.038}_{-0.037}$&\zuoneref&$0.179^{+0.038}_{-0.039}$\\
~~~~$u_{2,~z'}$\dotfill &$z$-band quadratic limb-darkening \dotfill &$0.268\pm0.044$&\zutworef&$0.288^{+0.046}_{-0.047}$\\
~~~~$u_{1,~\mathrm{TESS}}$\dotfill &TESS linear limb-darkening\dotfill &$0.267\pm0.016$&\tuoneref&$0.267^{+0.016}_{-0.015}$\\
~~~~$u_{2,~\mathrm{TESS}}$\dotfill &TESS quadratic limb-darkening\dotfill &$0.163^{+0.031}_{-0.030}$&\tutworef&$0.162^{+0.029}_{-0.031}$\\
\multicolumn{2}{l}{Doppler Tomography Parameters:}&&&\smallskip\\
~~~~$\sigma_{DT}$\dotfill &Doppler Tomography Error scaling \dotfill &$0.9933^{+0.0095}_{-0.0094}$&$0.9932^{+0.0095}_{-0.0094}$&$0.9941^{+0.0094}_{-0.0095}$\\
~~~~$\sigma_{DT}$\dotfill &Doppler Tomography Error scaling \dotfill &$0.9961^{+0.0096}_{-0.0094}$&$0.9963^{+0.0095}_{-0.0094}$&$0.9959^{+0.0097}_{-0.0098}$\\
~~~~$\sigma_{DT}$\dotfill &Doppler Tomography Error scaling \dotfill &$0.9667^{+0.0060}_{-0.0059}$&$0.9668^{+0.0061}_{-0.0060}$&$0.9670^{+0.0058}_{-0.0064}$\\
\multicolumn{5}{c}{Telescope Parameters}\\
\hline
\multicolumn{2}{l}{MINERVA T1:}&&&\smallskip\\
~~~~$\gamma_{\rm rel}$\dotfill &Relative RV Offset$^{5}$ [$\mathrm{m~s^{-1}}$]\dotfill &$-36^{+18}_{-17}$&$-38^{+18}_{-17}$&$-39^{+17}_{-16}$\\
~~~~$\sigma_{\rm J}$\dotfill &RV Jitter [$\mathrm{m~s^{-1}}$]\dotfill &$62^{+20}_{-19}$&$60\pm20$&$59\pm20$\\
~~~~$\sigma_{\rm J^2}$\dotfill &RV Jitter Variance \dotfill &$3900^{+2800}_{-2000}$&$3700^{+2800}_{-2000}$&$3600^{+2800}_{-2000}$\\
\multicolumn{2}{l}{MINERVA T2:}&&&\smallskip\\
~~~~$\gamma_{\rm rel}$\dotfill &Relative RV Offset$^{5}$ [$\mathrm{m~s^{-1}}$]\dotfill &$1\pm18$&$-1\pm18$&$0\pm19$\\
~~~~$\sigma_{\rm J}$\dotfill &RV Jitter [$\mathrm{m~s^{-1}}$]\dotfill &$45^{+23}_{-28}$&$47^{+22}_{-26}$&$46^{+22}_{-25}$\\
~~~~$\sigma_{\rm J^2}$\dotfill &RV Jitter Variance \dotfill &$2000^{+2600}_{-1800}$&$2200^{+2500}_{-1800}$&$2200^{+2500}_{-1700}$\\
\multicolumn{2}{l}{MINERVA T3:}&&&\smallskip\\
~~~~$\gamma_{\rm rel}$\dotfill &Relative RV Offset$^{5}$ [$\mathrm{m~s^{-1}}$]\dotfill &$46^{+20}_{-19}$&$44\pm20$&$44\pm19$\\
~~~~$\sigma_{\rm J}$\dotfill &RV Jitter [$\mathrm{m~s^{-1}}$]\dotfill &$77^{+22}_{-20}$&$77^{+22}_{-20}$&$78^{+22}_{-20}$\\
~~~~$\sigma_{\rm J^2}$\dotfill &RV Jitter Variance \dotfill &$6000^{+3900}_{-2700}$&$6000^{+3800}_{-2700}$&$6100^{+3800}_{-2800}$\\
\multicolumn{2}{l}{MINERVA T4:}&&&\smallskip\\
~~~~$\gamma_{\rm rel}$\dotfill &Relative RV Offset$^{5}$ [$\mathrm{m~s^{-1}}$]\dotfill &$-92\pm25$&$-93^{+25}_{-24}$&$-94^{+26}_{-23}$\\
~~~~$\sigma_{\rm J}$\dotfill &RV Jitter [$\mathrm{m~s^{-1}}$]\dotfill &$86^{+28}_{-25}$&$87^{+28}_{-25}$&$87^{+27}_{-25}$\\
~~~~$\sigma_{\rm J^2}$\dotfill &RV Jitter Variance \dotfill &$7400^{+5600}_{-3700}$&$7700^{+5700}_{-3800}$&$7600^{+5500}_{-3700}$\\
\multicolumn{2}{l}{SONG:}&&&\smallskip\\
~~~~$\gamma_{\rm rel,}$\dotfill &Relative RV Offset$^{5}$ [$\mathrm{m~s^{-1}}$]\dotfill &$411^{+19}_{-18}$&$408^{+19}_{-18}$&$408^{+19}_{-17}$\\
~~~~$\sigma_{\rm J,}$\dotfill &RV Jitter [$\mathrm{m~s^{-1}}$]\dotfill &$67.1^{+9.1}_{-8.1}$&$67.3^{+9.1}_{-8.1}$&$67.1^{+9.0}_{-8.1}$\\
~~~~$\sigma_{\rm J^2}$\dotfill &RV Jitter Variance \dotfill &$4500^{+1300}_{-1000}$&$4500^{+1300}_{-1000}$&$4500^{+1300}_{-1000}$\\
\multicolumn{2}{l}{TRES:}&&&\smallskip\\
~~~~$\gamma_{\rm rel}$\dotfill &Relative RV Offset$^{5}$ [$\mathrm{m~s^{-1}}$]\dotfill &$225^{+13}_{-12}$&$221^{+13}_{-12}$&$221^{+13}_{-11}$\\
~~~~$\sigma_{\rm J}$\dotfill &RV Jitter [$\mathrm{m~s^{-1}}$]\dotfill &$18^{+12}_{-19}$&$16^{+13}_{-16}$&$16^{+12}_{-16}$\\
~~~~$\sigma_{\rm J^2}$\dotfill &RV Jitter Variance \dotfill &$360^{+590}_{-370}$&$260^{+560}_{-340}$&$260^{+510}_{-340}$\\
\multicolumn{5}{c}{Transit Parameters$^\dagger$}\\
\hline
\multicolumn{2}{l}{KeplerCam UT 2019-03-30 (i')}&&&\smallskip\\
~~~~$\sigma^{2}$\dotfill &Added Variance \dotfill &$0.00000324^{+0.00000035}_{-0.00000031}$&$0.00000320^{+0.00000035}_{-0.00000031}$&$0.00000333^{+0.00000036}_{-0.00000031}$\\
~~~~$A_{\rm D}$\dotfill &Dilution from neighboring stars \dotfill &$0.0156^{+0.0020}_{-0.0018}$&&$0.077^{+0.019}_{-0.017}$\\
~~~~$F_{\rm 0}$\dotfill &Baseline flux \dotfill &$1.00064\pm0.00013$&$1.00074\pm0.00013$&$1.00027^{+0.00016}_{-0.00017}$\\
~~~~$C_{\rm 0}$\dotfill &Additive detrending coeff \dotfill &$-0.00078\pm0.00033$&$-0.00088\pm0.00032$&$-0.00038^{+0.00034}_{-0.00035}$\\
\multicolumn{2}{l}{ULMT UT 2019-03-30 (z')}&&&\smallskip\\
~~~~$\sigma^{2}$\dotfill &Added Variance \dotfill &$0.00000111^{+0.00000016}_{-0.00000014}$&$0.00000112^{+0.00000016}_{-0.00000014}$&$0.00000113^{+0.00000016}_{-0.00000015}$\\
~~~~$A_{\rm D}$\dotfill &Dilution from neighboring stars \dotfill &$0.0248^{+0.0032}_{-0.0030}$&&$0.093^{+0.019}_{-0.018}$\\
~~~~$F_{\rm 0}$\dotfill &Baseline flux \dotfill &$1.000157^{+0.000088}_{-0.000087}$&$1.000248^{+0.000087}_{-0.000086}$&$0.99990^{+0.00011}_{-0.00010}$\\
~~~~$C_{\rm 0}$\dotfill &Additive detrending coeff \dotfill &$-0.00082^{+0.00033}_{-0.00034}$&$-0.00105\pm0.00033$&$-0.00015^{+0.00040}_{-0.00037}$\\
~~~~$C_{\rm 1}$\dotfill &Additive detrending coeff \dotfill &$-0.00033\pm0.00023$&$-0.00043\pm0.00023$&$-0.00005^{+0.00026}_{-0.00024}$\\
\multicolumn{2}{l}{KeplerCam UT 2019-04-10 (i')}&&&\smallskip\\
~~~~$\sigma^{2}$\dotfill &Added Variance \dotfill &$0.00000829^{+0.0000011}_{-0.00000097}$&$0.00000830^{+0.0000011}_{-0.00000096}$&$0.00000832^{+0.0000011}_{-0.00000097}$\\
~~~~$A_{\rm D}$\dotfill &Dilution from neighboring stars \dotfill &$0.0158^{+0.0020}_{-0.0019}$&&$0.083^{+0.021}_{-0.019}$\\
~~~~$F_{\rm 0}$\dotfill &Baseline flux \dotfill &$1.00007\pm0.00025$&$1.00011\pm0.00025$&$0.99988^{+0.00025}_{-0.00024}$\\
~~~~$C_{\rm 0}$\dotfill &Additive detrending coeff \dotfill &$-0.00010^{+0.00062}_{-0.00063}$&$0.00006\pm0.00063$&$-0.00063^{+0.00064}_{-0.00065}$\\
\multicolumn{2}{l}{ULMT UT 2019-04-11 (z')}&&&\smallskip\\
~~~~$\sigma^{2}$\dotfill &Added Variance \dotfill &$0.00000680^{+0.00000085}_{-0.00000074}$&$0.00000679^{+0.00000085}_{-0.00000074}$&$0.00000687^{+0.00000084}_{-0.00000075}$\\
~~~~$A_{\rm D}$\dotfill &Dilution from neighboring stars \dotfill &$0.0246^{+0.0032}_{-0.0030}$&&$0.092^{+0.020}_{-0.018}$\\
~~~~$F_{\rm 0}$\dotfill &Baseline flux \dotfill &$0.99989\pm0.00020$&$0.99999\pm0.00020$&$0.99965^{+0.00022}_{-0.00023}$\\
\multicolumn{2}{l}{SOTES UT 2019-04-16 (R)}&&&\smallskip\\
~~~~$\sigma^{2}$\dotfill &Added Variance \dotfill &$-0.00000067^{+0.00000047}_{-0.00000039}$&$-0.00000062^{+0.00000049}_{-0.00000041}$&$-0.00000083^{+0.00000046}_{-0.00000039}$\\
~~~~$A_{\rm D}$\dotfill &Dilution from neighboring stars \dotfill &$0.0099^{+0.0013}_{-0.0012}$&&$0.073^{+0.019}_{-0.018}$\\
~~~~$F_{\rm 0}$\dotfill &Baseline flux \dotfill &$1.00067^{+0.00018}_{-0.00019}$&$1.00073^{+0.00018}_{-0.00019}$&$1.00036^{+0.00020}_{-0.00019}$\\
~~~~$C_{\rm 0}$\dotfill &Additive detrending coeff \dotfill &$-0.00039^{+0.00040}_{-0.00039}$&$-0.00041\pm0.00040$&$-0.00024^{+0.00039}_{-0.00040}$\\
\multicolumn{2}{l}{CROW UT 2019-04-27 (i')}&&&\smallskip\\
~~~~$\sigma^{2}$\dotfill &Added Variance \dotfill &$0.0000318^{+0.0000050}_{-0.0000042}$&$0.0000318^{+0.0000049}_{-0.0000042}$&$0.0000319^{+0.0000046}_{-0.0000041}$\\
~~~~$A_{\rm D}$\dotfill &Dilution from neighboring stars \dotfill &$0.0157^{+0.0020}_{-0.0018}$&&$0.081^{+0.020}_{-0.018}$\\
~~~~$F_{\rm 0}$\dotfill &Baseline flux \dotfill &$2.93835\pm0.00053$&$2.93861\pm0.00054$&$2.93738^{+0.00058}_{-0.00062}$\\
~~~~$C_{\rm 0}$\dotfill &Additive detrending coeff \dotfill &$0.00323^{+0.00043}_{-0.00042}$&$0.00334^{+0.00042}_{-0.00041}$&$0.00278^{+0.00043}_{-0.00044}$\\
\multicolumn{2}{l}{LCO$_{TFN}$ UT 2019-04-27 (z')}&&&\smallskip\\
~~~~$\sigma^{2}$\dotfill &Added Variance \dotfill &$0.0000333^{+0.0000034}_{-0.0000031}$&$0.0000331^{+0.0000033}_{-0.0000030}$&$0.0000333^{+0.0000033}_{-0.0000029}$\\
~~~~$A_{\rm D}$\dotfill &Dilution from neighboring stars \dotfill &$0.0245^{+0.0032}_{-0.0029}$&&$0.093^{+0.020}_{-0.018}$\\
~~~~$F_{\rm 0}$\dotfill &Baseline flux \dotfill &$1.00561^{+0.00036}_{-0.00035}$&$1.00574^{+0.00035}_{-0.00036}$&$1.00524^{+0.00036}_{-0.00034}$\\
~~~~$C_{\rm 0}$\dotfill &Additive detrending coeff \dotfill &$0.00038^{+0.00081}_{-0.00080}$&$0.00050^{+0.00080}_{-0.00081}$&$0.00017^{+0.00081}_{-0.00082}$\\
\multicolumn{2}{l}{KAO UT 2019-05-03 (z')}&&&\smallskip\\
~~~~$\sigma^{2}$\dotfill &Added Variance \dotfill &$0.00000563^{+0.00000053}_{-0.00000049}$&$0.00000572^{+0.00000054}_{-0.00000049}$&$0.00000546^{+0.00000054}_{-0.00000050}$\\
~~~~$A_{\rm D}$\dotfill &Dilution from neighboring stars \dotfill &$0.0251^{+0.0033}_{-0.0031}$&&$0.099^{+0.022}_{-0.020}$\\
~~~~$F_{\rm 0}$\dotfill &Baseline flux \dotfill &$1.00135\pm0.00014$&$1.00147\pm0.00014$&$1.00103^{+0.00016}_{-0.00017}$\\
~~~~$C_{\rm 0}$\dotfill &Additive detrending coeff \dotfill &$0.00137\pm0.00031$&$0.00128^{+0.00031}_{-0.00032}$&$0.00171^{+0.00033}_{-0.00031}$\\
\multicolumn{2}{l}{KAP UT 2019-05-03 (z')}&&&\smallskip\\
~~~~$\sigma^{2}$\dotfill &Added Variance \dotfill &$0.0000258^{+0.0000027}_{-0.0000024}$&$0.0000257^{+0.0000028}_{-0.0000024}$&$0.0000259^{+0.0000028}_{-0.0000025}$\\
~~~~$A_{\rm D}$\dotfill &Dilution from neighboring stars \dotfill &$0.0245^{+0.0032}_{-0.0030}$&&$0.093^{+0.020}_{-0.019}$\\
~~~~$F_{\rm 0}$\dotfill &Baseline flux \dotfill &$0.99987\pm0.00033$&$1.00000\pm0.00033$&$0.99952^{+0.00034}_{-0.00036}$\\
~~~~$C_{\rm 0}$\dotfill &Additive detrending coeff \dotfill &$-0.00070\pm0.00063$&$-0.00080^{+0.00062}_{-0.00061}$&$-0.00038^{+0.00062}_{-0.00063}$\\
\multicolumn{2}{l}{TESS UT 2019-07-18 (TESS)}&&&\smallskip\\
~~~~$\sigma^{2}$\dotfill &Added Variance \dotfill &$\left(7.40\pm0.22\right)\times10^{-8}$&$\left(7.40\pm0.22\right)\times10^{-8}$&$\left(7.42\pm0.22\right)\times10^{-8}$\\
~~~~$F_{\rm 0}$\dotfill &Baseline flux \dotfill &$1.0000119\pm0.0000034$&$1.0000120\pm0.0000034$&$1.0000117^{+0.0000036}_{-0.0000034}$\\
\multicolumn{2}{l}{TESS UT 2019-12-25 (TESS)}&&&\smallskip\\
~~~~$\sigma^{2}$\dotfill &Added Variance \dotfill &$\left(-1.39\pm0.15\right)\times10^{-8}$&$\left(-1.40\pm0.15\right)\times10^{-8}$&$\left(-1.39^{+0.15}_{-0.14}\right)\times10^{-8}$\\
~~~~$F_{\rm 0}$\dotfill &Baseline flux \dotfill &$1.0000055^{+0.0000028}_{-0.0000029}$&$1.0000055\pm0.0000028$&$1.0000053^{+0.0000030}_{-0.0000027}$\\
\multicolumn{2}{l}{TESS UT 2020-01-21 (TESS)}&&&\smallskip\\
~~~~$\sigma^{2}$\dotfill &Added Variance \dotfill &$\left(2.74\pm0.16\right)\times10^{-8}$&$\left(2.74\pm0.16\right)\times10^{-8}$&$\left(2.74\pm0.15\right)\times10^{-8}$\\
~~~~$F_{\rm 0}$\dotfill &Baseline flux \dotfill &$1.0000058^{+0.0000028}_{-0.0000029}$&$1.0000058\pm0.0000029$&$1.0000057^{+0.0000029}_{-0.0000030}$\\
\multicolumn{2}{l}{TESS UT 2021-06-25 (TESS)}&&&\smallskip\\
~~~~$\sigma^{2}$\dotfill &Added Variance \dotfill &$\left(21.29\pm0.33\right)\times10^{-8}$&$\left(21.30^{+0.34}_{-0.33}\right)\times10^{-8}$&$\left(21.29^{+0.34}_{-0.33}\right)\times10^{-8}$\\
~~~~$F_{\rm 0}$\dotfill &Baseline flux \dotfill &$0.9999702^{+0.0000042}_{-0.0000041}$&$0.9999701\pm0.0000041$&$0.9999701^{+0.0000041}_{-0.0000040}$\\
\multicolumn{2}{l}{TESS UT 2021-07-24 (TESS)}&&&\smallskip\\
~~~~$\sigma^{2}$\dotfill &Added Variance \dotfill &$\left(6.55\pm0.20\right)\times10^{-8}$&$\left(6.54\pm0.20\right)\times10^{-8}$&$\left(6.56^{+0.20}_{-0.21}\right)\times10^{-8}$\\
~~~~$F_{\rm 0}$\dotfill &Baseline flux \dotfill &$1.0000027\pm0.0000032$&$1.0000028\pm0.0000032$&$1.0000029^{+0.0000033}_{-0.0000032}$\\
\multicolumn{2}{l}{TESS UT 2021-12-31 (TESS)}&&&\smallskip\\
~~~~$\sigma^{2}$\dotfill &Added Variance \dotfill &$\left(4.54\pm0.17\right)\times10^{-8}$&$\left(4.54\pm0.17\right)\times10^{-8}$&$\left(4.55^{+0.16}_{-0.17}\right)\times10^{-8}$\\
~~~~$F_{\rm 0}$\dotfill &Baseline flux \dotfill &$1.0000054^{+0.0000030}_{-0.0000031}$&$1.0000053\pm0.0000030$&$1.0000053^{+0.0000031}_{-0.0000030}$\\
\multicolumn{2}{l}{TESS UT 2022-01-30 (TESS)}&&&\smallskip\\
~~~~$\sigma^{2}$\dotfill &Added Variance \dotfill &$\left(4.52\pm0.18\right)\times10^{-8}$&$\left(4.52\pm0.18\right)\times10^{-8}$&$\left(4.53\pm0.18\right)\times10^{-8}$\\
~~~~$F_{\rm 0}$\dotfill &Baseline flux \dotfill &$1.0000057\pm0.0000032$&$1.0000057\pm0.0000032$&$1.0000059\pm0.0000033$
\enddata
\tablenotetext{}{See Table 3 in \citet{Eastman:2019} for a detailed description of all parameters}
\tablenotetext{\dagger}{Instruments with missing dilution or additive detrending coefficients did not require them}
\tablenotetext{1}{-2450000; time of conjunction is commonly reported as the "transit time"}
\tablenotetext{2}{-2450000; time of minimum projected separation is a more correct "transit time"}
\tablenotetext{3}{-2450000; optimal time of conjunction minimizes the covariance between $T_C$ and Period}
\tablenotetext{4}{Assumes no albedo and perfect redistribution}
\tablenotetext{5}{Reference epoch = 2458984.524152}
\end{deluxetable*}
~
~

~
~
~
\vspace{2cm}
~
~
~

\section{Evidence for Additional Bodies in the KELT-24 System} \label{sec:other_planets}

\begin{figure*}
    \centering
    \includegraphics[width=\linewidth]{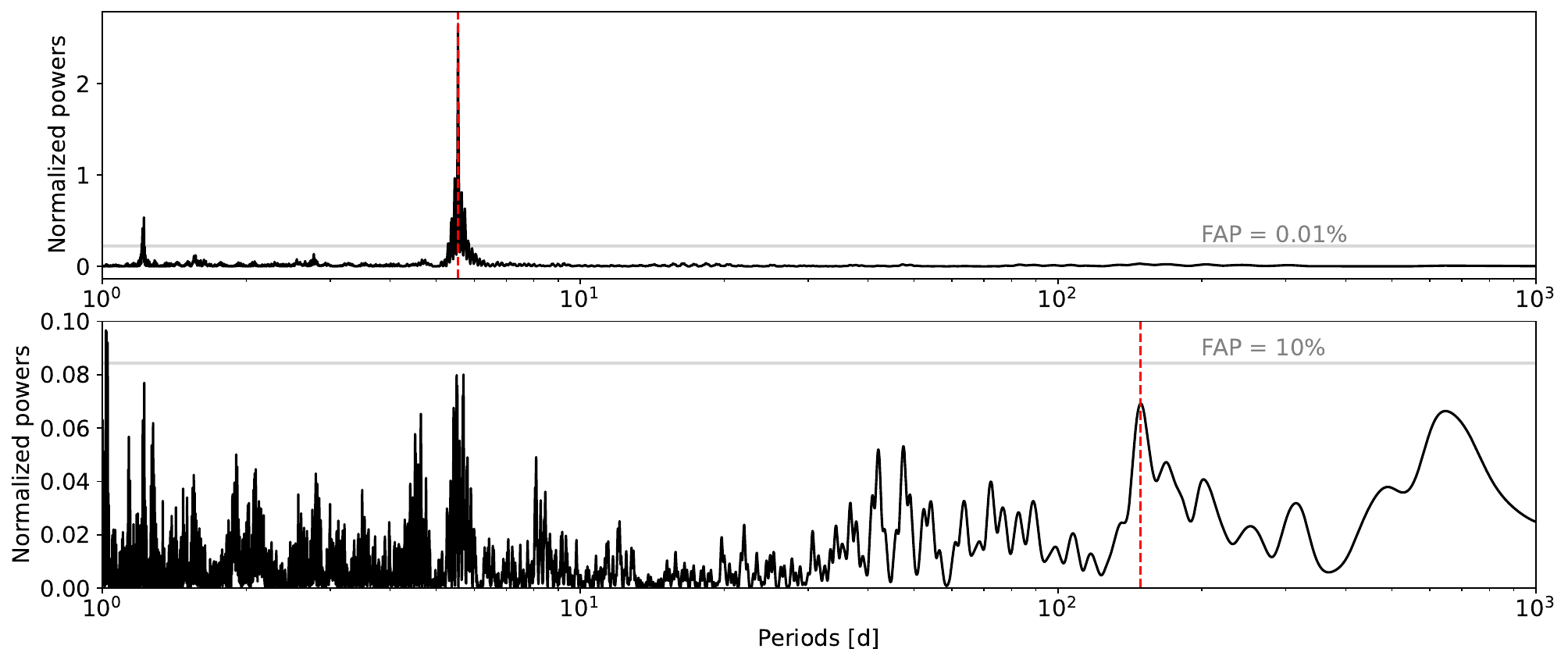}
    \caption{Top panel: Lomb-scargle \citep{lomb, scargle} periodogram in the RV data presented in this analysis. The $\approx5.55$~d period of KELT-24~b is strongly evident. Bottom panel: Lomb-scargle periodogram of the residual RVs. We find weak evidence for periodicity at 149.2~d.}
    \label{fig:rv_periodogram}
\end{figure*}

Here, we aim to search for and place constraints on additional bodies in the system. To search for the tentative KELT-24~c, we apply our RV modeling toolkit described in \cite{2021arXiv210913996C}. We apply a two-planet model to our RVs and generate a log-likelihood for a wide range in fixed periods for $P_\mathrm{c}$, fitting only for $T_\mathrm{Cc}$ and $K_\mathrm{c}$. Due to its large contribution to the RV variations, we include KELT-24~b as well but using Gaussian priors for the orbital parameters informed from the global fit results in Table \ref{tab:planet_posteriors}. We also include the relative instrument dependent offsets and additional radial velocity jitter noise terms in the model. We find weak evidence for an additional companion with $P\sim150$~d in our RV measurements (see Figure \ref{fig:rv_periodogram}).

\begin{figure}[h]
    \centering
    \includegraphics[width=\linewidth]{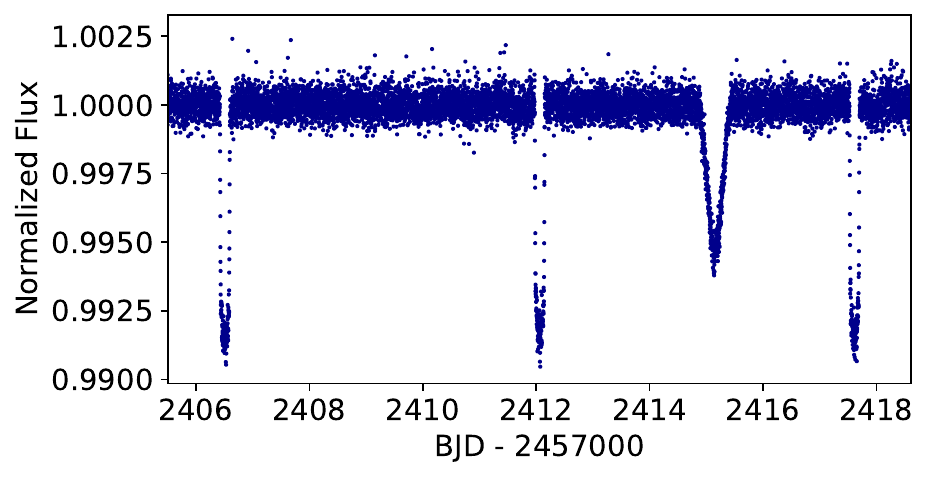}
    \caption{Snippet from TESS Sector 40 highlighting the single transit ($\mathrm{BJD}-2457000\approx2415$) in the KELT-24 (TIC 349827430) field of view. KELT-24~b can be seen transiting three times ($\mathrm{BJD}-2457000\approx2406.5,~2412.0,~2417.5$).}
    \label{fig:tess_single_transit}
\end{figure}

\begin{figure}[h]
    \centering
    \includegraphics[width=\linewidth]{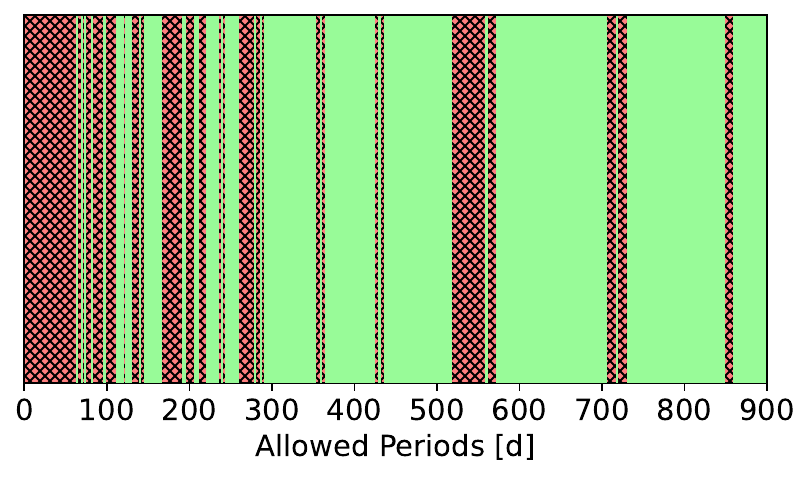}
    \caption{Allowed periods for a bound companion as the source of the single transit observed in TESS photometry are shaded in green, while possible periods forbidden are hatched and shaded in red.}
    \label{fig:allowed_periods}
\end{figure}

\begin{figure}[h]
    \centering
    \includegraphics[width=\linewidth]{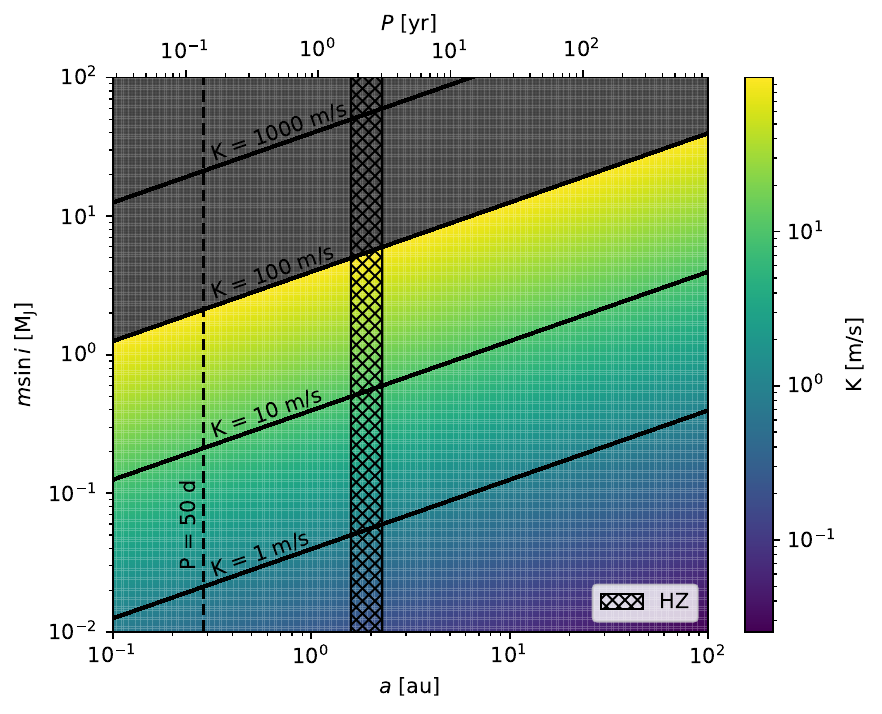}
    \caption{Masses and periods of potential planetary counterparts to KELT-24~b that are not restricted by available RVs and photometry. Parameter space restricted by photometric and spectroscopic observations is shown in gray. Our supposed RV sensitivity limit of $K=100~\mathrm{m~s^{-1}}$ is conservative and based on the standard error of our current observations. We place a somewhat arbitrary $P=50~\mathrm{d}$ threshold as a conservative minimum period for an additional body excluded via absence in TESS photometry, but note that slightly misaligned planets with shorter periods cannot be ruled out. The estimated habitable zone (HZ) is shown as the narrow hatched region.}
    \label{fig:planet_sensitivity}
\end{figure}

The discovery of KELT-24~b by R2019 and H2019 preceded the first release of TESS, and so no TESS photometry was included in the original analysis. \cite{2020AcA....70..181M} later noted that KELT-24 was observed in Sectors 14, 20, and 21. We also analyzed photometry from TESS, which offers a long baseline search for companion planets in a system already known to host one transiting planet in KELT-24~b. While the $\approx$5.55 day period of KELT-24~b was evident across all sectors listed in Table \ref{tab:tess_sec}, one additional deep single transit ($t_\mathrm{dur}\approx12~\mathrm{hr}$) is observed in Sector 40 at $\mathrm{JD}=2459414$, and is highlighted in Figure \ref{fig:tess_single_transit}. This signal is not found elsewhere in the TESS observations, but could be evidence of a hierarchical, gravitationally-bound body in the KELT-24 system. Since we do not observe any portion of this transit-like event elsewhere in the TESS photometry, we can assume that it was only captured once and use that information to restrict regions of period space for the transiting body. The allowed periods derived from our TESS observing baseline are depicted in Figure \ref{fig:allowed_periods}. Conservatively, we can assert that our RV observations rule out any long-period planet with $K>100~\mathrm{m~s^{-1}}$, and assume that any such companion would have $P>50~\mathrm{d}$. The parameter space therefore allowed for possible planets that our analyses would be insensitive to are shown in Figure \ref{fig:planet_sensitivity}. A Jupiter-mass planet on a multi-year orbit could easily be hidden beneath the noise floor of our RV observations. Notably, many lesser-mass planets on the $P\approx150~\mathrm{d}$ orbit discussed above cannot be faithfully excluded.

As discussed in Section \ref{sec:imaging}, KELT-24 has a likely bound stellar companion, nominally dubbed KELT-24B, $\approx2\arcsec$ away. Though even Gaia only resolves it as a single source, it is possible that KELT-24B is an unresolved, eclipsing binary star system, KELT-24BC. As previously outlined, KELT-24B's placement in the HR diagram is relatively bright for stars of similar colors. Specifically, given its separation from the MIST evolutionary track shown in Figure \ref{fig:HR} for an age and metallicity consistent with those derived in Section \ref{sec:global_fit}, KELT-24B is better described as an unresolved multiple system, likely a binary pair of two M-type stars, each with $M_*\approx0.50~\mathrm{M_\odot}$ (see Table \ref{tab:KELT-24.3stars.2.}). Under the assumption that this source may indeed be an unresolved binary, KELT-24BC, we maintain that it could be also be the origin of the single transit. The single transit depth is $\sim0.5\%$ of the baseline flux, and considering the $\sim1-3\%$ flux contribution from the background companion in the TESS bandpass, an unresolved eclipsing binary like the one our evolutionary models suggest could indeed manifest as a transit with the properties observed in Sector 40 of TESS and shown in Figure \ref{fig:tess_single_transit}. Given the high RUWE for KELT-24B, the depth of the single eclipse, and the fact that KELT-24B appears better described as two stars, it seems likely that the eclipse event did not result from a body transiting KELT-24, but rather from an eclipse in the binary system KELT-24BC. We note, however, that the supposed multiple system shares close proximity with its much brighter companion, and so KELT-24B's high RUWE may be a result of blending from KELT-24 rather than an indication that the source is a hierarchical system. We attempted to take an isolated spectrum of nearby companion KELT-24B with TRES, but were limited by the 2.3$\arcsec$ fiber and therefore unable to resolve it. High-resolution spectroscopy with AO, or from a site with good seeing, may be capable of determining the true nature of this source. For example, \cite{2016ApJ...817...80D} reported several RVs of both components of Kepler-444 BC ($\rho=1.8''$) using Keck/HIRES \citep{1994SPIE.2198..362V} on nights with very good seeing. A longer RV-baseline or lucky transit photometry is needed to confirm the nature of the single transit found in Sector 40.

The SPOC transit search is configured for TESS to require a minimum of two transit events to define a Threshold Crossing Event (TCE). The SPOC data validation component \citep{2018PASP..130f4502T, 2019PASP..131b4506L} was provided with the Sector 40 light curve and calibrated pixel data, and run in a supplemental science mode with an ephemeris including only the single transit. The resulting difference image centroid offset excluded all TIC v8 objects \citep{2019AJ....158..138S} other than KELT-24 and KELT-24B as potential sources of the transit. The centroid offset located the transit source within $5.15''$ ($2.06~\sigma$) of KELT-24 and $3.07''$ ($1.23~\sigma$) of KELT-24B. While not definitive, the centroid offset suggests that the single Sector 40 transit is more likely to be an eclipse of the companion than a transit of KELT-24, supporting the claim that KELT-24B is indeed an unresolved stellar multiple. Interestingly, if the single transit is ultimately linked to KELT-24B, it would imply that the KELT-24 system contains an orbiting pair of two edge-on systems (KELT-24/KELT-24~b and KELT-24BC, in this case). Yet, the orbital inclination between KELT-24 and the presumed eclipsing binary estimated in Section \ref{subsec:gaia} of $i=147.57^{+14.57\circ}_{-13.47}$ suggests that the two edge-on systems are relatively inclined (by $\approx2\sigma$). While the Gaia proper motion measurements (and therefore the orbit's inclination) of the supposed eclipsing binary should be taken with caution, given that they are derived assuming a single source, the relative inclinations of the four-body configuration between KELT-24/KELT-24~b and KELT-24BC would serve as a sandbox for tests of planetary and solar system formation and evolution theories.

With mounting evidence that the bound source to KELT-24 is an unresolved (likely eclipsing) pair, we considered a three-star fit (KELT-24, KELT-24B, KELT-24C) to quantify the biases that our two-source fits from Section \ref{sec:global_fit} may have incurred. To execute this, we model the stellar parameters of KELT-24 following the method outlined in Section \ref{subsec:mascara_reproduce} assuming the SED is blended by two nearby sources blended as one. Since this method of modeling first derives only the stellar parameters, we can compare the results of the primary between both sets of posteriors to determine whether or not fitting for an additional faint source induces and noticeable changes to KELT-24. The results of this star-only fit are provided in Table \ref{tab:KELT-24.3stars.2.}. We observe that KELT-24's derived stellar parameters from both Table \ref{tab:derived_stellar_params} (Section \ref{subsec:mascara_reproduce}) and Table \ref{tab:KELT-24.3stars.2.} (column A) are consistent, suggesting that the inclusion of a third body in our overall fits contribute minimally to the derived parameters of KELT-24 and KELT-24~b. Additionally, the sum of the individual luminosities from this fit are consistent with the computed luminosity of the second source from our two-component fit used to determine the values published in Table \ref{tab:derived_stellar_params}. This, combined with the fact that each individual stellar temperature for KELT-24B and KELT-24C here is in line with the estimated value for that of our secondary from the two-source fit, is indicative that the three-star fit is in good agreement with the available data we have for the KELT-24 system. For this reason, we maintain that our three fits presented in Section \ref{sec:global_fit} are well-motivated and unbiased. Our three-star fit finds the eclipsing binary to consist of two half-solar mass stars. Assuming the masses and radii for the presumed eclipsing binary pair reported from our fit, we estimate a grazing probability of $\approx0.01$ for orbital periods near 100~d, and $\approx0.001$ for orbital periods around 10~yr. We note that the implied total mass ($1.028~\mathrm{M_\odot}$) of the likely unresolved pair is marginally higher than the one assumed for our orbital analysis of the two resolved sources presented in Section \ref{subsec:gaia}, but that this will result in a minimal impact of the estimated orbit given the projected uncertainties.

\startlongtable
\begin{deluxetable*}{lcccc}
\tablecaption{Median values and 68\% confidence interval for KELT-24 stellar system assuming three stars (KELT-24A and KELT-24BC)}
\tablehead{\colhead{~~~Parameter} & \colhead{Description} & \multicolumn{3}{c}{Values}}
\startdata
\multicolumn{2}{l}{}&A&B&C\smallskip\\
~~~~$M_*$\dotfill &Mass [\msun]\dotfill &$1.309^{+0.052}_{-0.18}$&$0.517^{+0.081}_{-0.11}$&$0.511^{+0.066}_{-0.089}$\\
~~~~$R_*$\dotfill &Radius [\rsun]\dotfill &$1.513^{+0.035}_{-0.031}$&$0.483^{+0.078}_{-0.093}$&$0.487^{+0.053}_{-0.090}$\\
~~~~$R_{*,\mathrm{SED}}$\dotfill &Radius$^{1}$ [\rsun]\dotfill &$1.464^{+0.018}_{-0.013}$&$0.485^{+0.084}_{-0.098}$&$0.497^{+0.059}_{-0.11}$\\
~~~~$L_*$\dotfill &Luminosity [\lsun]\dotfill &$3.54^{+0.15}_{-0.16}$&$0.037\pm0.019$&$0.038^{+0.015}_{-0.018}$\\
~~~~$F_{Bol}$\dotfill &Bolometric Flux [cgs]\dotfill &$0.00000001212^{+0.00000000047}_{-0.00000000049}$&$0.000000000128^{+0.000000000063}_{-0.000000000066}$&$0.000000000132^{+0.000000000051}_{-0.000000000061}$\\
~~~~$\rho_*$\dotfill &Density [cgs]\dotfill &$0.525^{+0.050}_{-0.068}$&$6.4^{+3.3}_{-1.7}$&$6.2^{+3.3}_{-1.2}$\\
~~~~$\log{g}$\dotfill &Surface gravity [cgs]\dotfill &$4.192^{+0.028}_{-0.066}$&$4.780^{+0.084}_{-0.067}$&$4.772^{+0.093}_{-0.047}$\\
~~~~$T_{\rm eff}$\dotfill &Effective Temperature [K]\dotfill &$6437^{+57}_{-64}$&$3630^{+120}_{-190}$&$3640^{+150}_{-240}$\\
~~~~$T_{\rm eff,SED}$\dotfill &Effective Temperature$^{1}$ [K]\dotfill &$6548^{+82}_{-92}$&$3600^{+140}_{-160}$&$3600^{+120}_{-190}$\\
~~~~$[{\rm Fe/H}]$\dotfill &Metallicity [dex]\dotfill &$0.168^{+0.068}_{-0.058}$&$0.297^{+0.083}_{-0.071}$&$0.308^{+0.098}_{-0.069}$\\
~~~~$[{\rm Fe/H}]_{0}$\dotfill &Initial Metallicity$^{2}$ \dotfill &$0.255^{+0.055}_{-0.059}$&$0.255^{+0.055}_{-0.059}$&$0.255^{+0.055}_{-0.059}$\\
~~~~$Age$\dotfill &Age [Gyr]\dotfill &$2.81^{+4.0}_{-0.79}$&$2.81^{+4.1}_{-0.79}$&$2.81^{+4.0}_{-0.79}$\\
~~~~$EEP$\dotfill &Equal Evolutionary Phase$^{3}$ \dotfill &$362^{+77}_{-16}$&$278^{+28}_{-13}$&$274.7^{+25}_{-7.7}$\\
~~~~$A_V$\dotfill &V-band extinction [mag]\dotfill &$0.074^{+0.040}_{-0.046}$&$0.074^{+0.040}_{-0.046}$&$0.074^{+0.040}_{-0.046}$\\
~~~~$\sigma_{SED}$\dotfill &SED photometry error scaling \dotfill &$1.13^{+1.2}_{-0.25}$&$26^{+24}_{-21}$&$17^{+19}_{-13}$\\
\enddata
\label{tab:KELT-24.3stars.2.}
\tablenotetext{}{See Table 3 in \citet{Eastman:2019} for a detailed description of all parameters}
\tablenotetext{1}{This value ignores the systematic error and is for reference only}
\tablenotetext{2}{The metallicity of the star at birth}
\tablenotetext{3}{Corresponds to static points in a star's evolutionary history. See \S2 in \citet{MIST0}.}
\end{deluxetable*}

We attempted to search for an unbound background star in a chance alignment behind KELT-24, one that may even bias KELT-24's light curves or SED. Since KELT-24 has a significant proper motion of $\sim65$~mas~yr$^{-1}$, we first examined an image from 1953 taken by the Palomar Observatory Sky Survey (POSS; \citealt{1952PASP...64..275H, 1963bad..book..481M}, but unfortunately the resolution was too poor to determine whether any background components were present. While we are unable to confidently reject the existence of a background or unresolved companion to KELT-24, its reported $\mathrm{RUWE}=1.014$ suggests that if one were to be present, it would be unlikely to be bright enough to introduce noticeable complications. The lack of evidence for any nearby stellar counterparts to KELT-24 is compounded by the absence of any sources observed above the contrast detection limits that can be placed via the non-redundant masking AO image from 2019 introduced in Section \ref{sec:imaging}. The contrast detection limits for sources around KELT-24 are presented in Table \ref{tab:contrastlimits}. Stars bright enough to contaminate the photometry would likely also be apparent in the RVs, but we see no evidence for a significant stellar-induced RV trend. 

Our efforts to search for additional components in the KELT-24 system reveal that, while there may be no meaningful unseen star blended with KELT-24, there is a potential body in the system present in the form of the single transit described above. This single transit could be the imprint of either a long-period planet orbiting KELT-24 or a detection of an eclipsing binary: KELT-24BC. If the latter is indeed the source of the transit, it would not only explain the high RUWE reported by Gaia, but may also serve as an explanation to the discrepancies observed between R2019 and H2019, as well as our reanalyses presented here.

\startlongtable
\begin{deluxetable}{l|c||ccc}
\tablecaption{KELT-24 contrast limits from Keck/NIRC2}
\tablehead{\colhead{} & \colhead{Filter} & \colhead{Br$\gamma$} &  \colhead{$K_\mathrm{p}$} & \colhead{$K_\mathrm{p}$+C06}
\\
\colhead{} & \colhead{Epoch} & \colhead{2019.3596} & \colhead{2019.4389} & \colhead{2019.4389}}
\startdata
\multirow{9}*{\rotatebox{90}{Contrast $\Delta m$ [mag] at $\rho=$ (mas)}} & 150 & 4.5 & 4.1 & ...  \\
& 200 & 4.9 & 5.3 & ...  \\
& 250 & 5.8 & 5.4 & ...  \\
& 300 & 5.7 & 6.0 & ...  \\
& 400 & 6.4 & 6.9 & 7.2  \\
& 500 & 6.9 & 6.8 & 7.2  \\
& 700 & 7.1 & 7.7 & 8.3  \\
& 1000 & 7.3 & 9.0 & 9.8  \\
& 1500 & 7.3 & 9.6 & 10.1  
\enddata
\tablenotetext{}{}
\label{tab:contrastlimits}
\end{deluxetable}

Lastly, we identify an M dwarf (Gaia DR3 1076970372392617472, 2MASS J10472835+7139298) 48'' away from KELT-24 that has a proper motion vector highly consistent with being bound to the stellar system at the focus of this paper. While its parallax ($\pi=4.9821\pm0.1071$~mas) suggests that the source is far more distant than the KELT-24 system, it is possible that the star's faintness ($G=18.06$) makes the measurement accuracy unreliable. The expected enhancements in astrometric precision from Gaia DR4, for example, may be needed to determine whether the star is indeed a part of the KELT-24 system, or whether its proper motion is simply coincidental to the two sources discussed here. In any case, the separation of this star in terms of both distance and brightness from the KELT-24 system imply that it would not impact the dynamics or photometry considered in our analyses here.



\section{Discussion} \label{sec:discussion}

\begin{figure}[h]
    \centering
    \includegraphics[width=\linewidth]{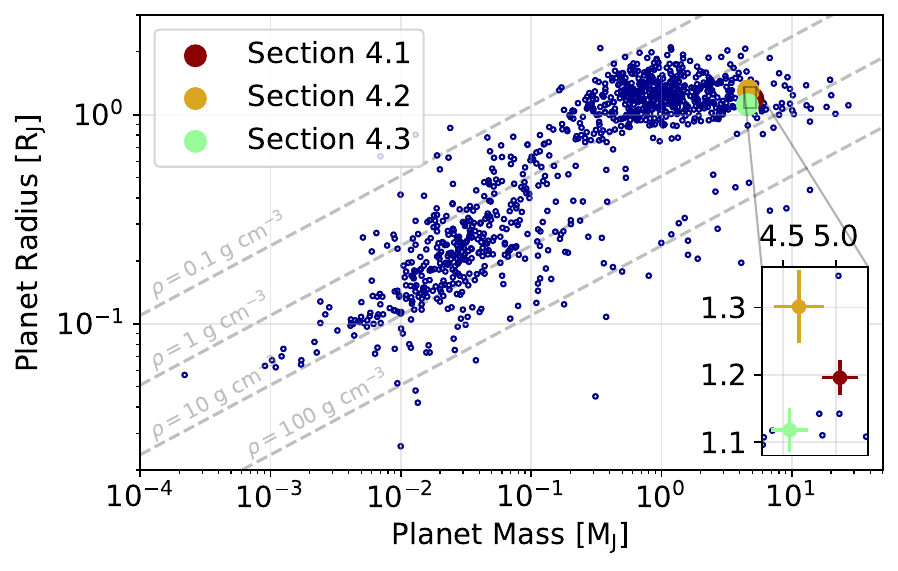}
    \caption{Mass-radius diagram of all known exoplanets with reported values for both parameters. The mass and radius of KELT-24~b determined by each of our individual global analyses is drawn in for reference. Density lines are also included. While there is good agreement between our three results here regarding the general nature of KELT-24~b as a large, dense planet, their internal discrepancies are apparent.}
    \label{fig:mass_radius}
\end{figure}

While the discrepancies in the derived physical parameeters for the KELT-24 system observed in Tables \ref{tab:derived_stellar_params} and \ref{tab:planet_posteriors} cannot presently be bridged, KELT-24~b is indisputably a well-aligned, transiting hot Jupiter orbiting a nearby, bright, main sequence star. Highlighted in Figure \ref{fig:mass_radius}, KELT-24~b is among the largest of known exoplanets in terms of both mass and physical size. With $\rho=3.90^{+0.19}_{-0.20}\mathrm{~g~cm^{-3}}$, it is also one of the densest giant exoplanets. Hereafter, we propagate the maximum likelihood parameters for each of the three fits outlined in our work. As is in Tables \ref{tab:derived_stellar_params} and \ref{tab:planet_posteriors}, our favored set of posterior solutions are denoted in green. The best-fit values from this solution set are those referenced in the text that follows.

\begin{figure}[h]
    \centering
    \includegraphics[width=\linewidth]{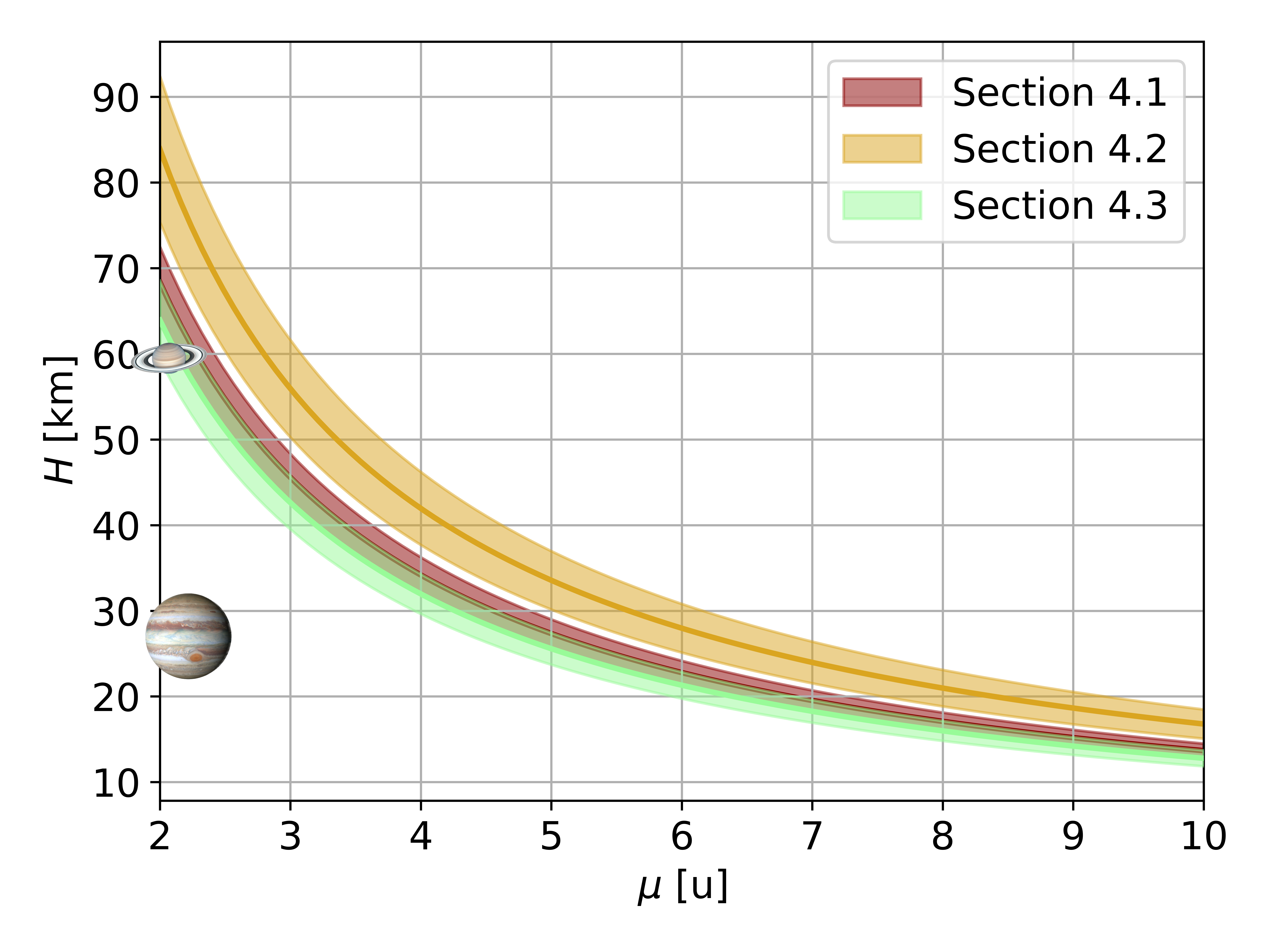}
    \caption{Scale height $H$ as a function of mean molecular weight $\mu$ using the best-fit properties for KELT-24~b derived in by each of our analyses Section \ref{sec:global_fit}. The shaded region is the 1$\sigma$ deviation in $H$ created by drawing from posteriors in $T_\mathrm{eq}$, $R_\mathrm{p}$, and $M_\mathrm{p}$. The mean molecular weights and scale heights of Jupiter and Saturn are shown for reference.}
    \label{fig:scale_height}
\end{figure}

KELT-24 is the brightest star known to host a transiting, short-period planet with a mass $>4~\mathrm{M_J}$, and is therefore an exciting prospect for atmospheric characterization. With its large size ($M_\mathrm{p}=4.59\pm0.18~\mathrm{M_J}$, $R_\mathrm{p}=1.134^{+0.027}_{-0.026}~\mathrm{R_J}$) and hot equilibrium temperature ($T_\mathrm{eq}=1463\pm10~\mathrm{K}$), we can estimate KELT-24~b's scale height $H$ as

\begin{equation} \label{eq:scale_height}
    H=\frac{k_\mathrm{B}T_\mathrm{eq}R_\mathrm{p}^2}{G\mu M_\mathrm{p}}
\end{equation}

where $k_\mathrm{B}$ is the Boltzmann constant, $G$ is the gravitational constant, and $\mu$ is the mean molecular weight of the atmosphere in atomic mass units. For an array of possible values for $\mu$ consistent with predictions for gas giants like KELT-24~b \citep[often $\approx2-3$, but rarely more than 10; e.g.,][]{mmw_1, mmw_2, mmw_3}, we determine the scale height for KELT-24~b to be of order $\gtrsim10$~km, as shown in Figure \ref{fig:scale_height}.

\begin{figure}[h]
    \centering
    \includegraphics[width=\linewidth]{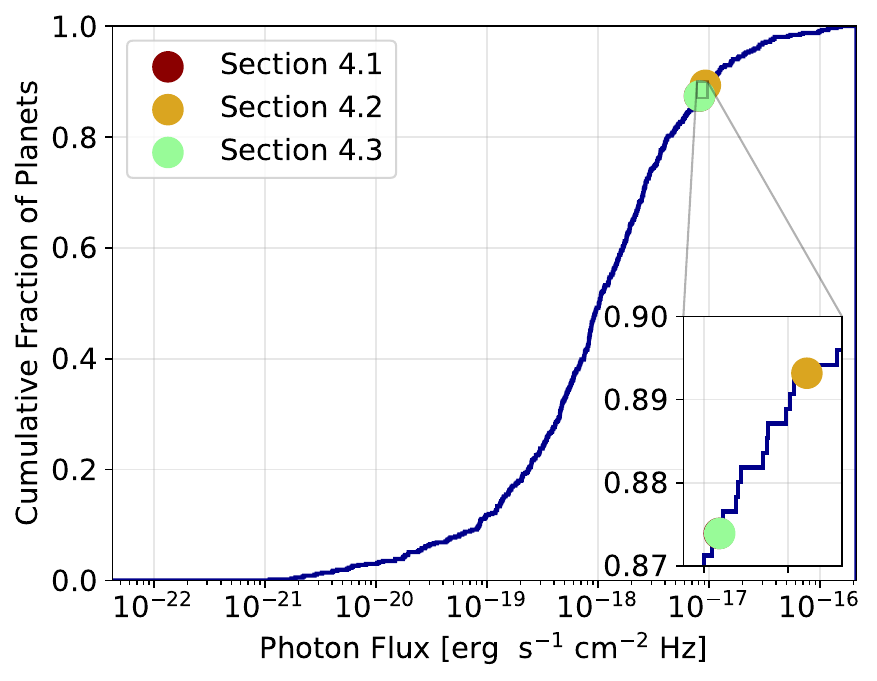}
    \caption{Expected flux through the atmosphere of all known transiting exoplanets. Each of our three derived solutions for KELT-24~b are included here, and in any case it is estimated to be in the top 13\% of all known transiting planets in terms of stellar flux passing through its atmosphere. We note that this derived flux is nearly identical for Sections \ref{subsec:rodriguez_reproduce} and \ref{subsec:no_sed}, and so they appear to overlap here.}
    \label{fig:atmospheric_flux}
\end{figure}

Considering KELT-24~b's $\gtrsim10$~km atmospheric scale height, as well as the proximity to its very bright ($V=8.3$) host star, few transiting planets are observed to have as many photons pass through their atmospheres (see Figure \ref{fig:atmospheric_flux}). It is therefore an excellent candidate for transmission spectroscopy studies via JWST \citep{jwst}. \cite{2018PASP..130k4401K} developed a metric (transmission spectroscopy metric; TSM) for quantifying the relative signal to noise that would be generated from a JWST transmission spectrum, and we find that KELT-24~b has $\mathrm{TSM}\approx41$, which is among the highest of all dense $\left(\rho>1~\mathrm{g~cm^{-3}}\right)$, giant $\left(R_\mathrm{p}>1~\mathrm{R_J}\right)$ exoplanets. KELT-24~b's short period ($P\approx5.55$~d) also makes it amenable to scheduling follow-up observations.

\begin{figure}[h]
    \centering
    \includegraphics[width=\linewidth]{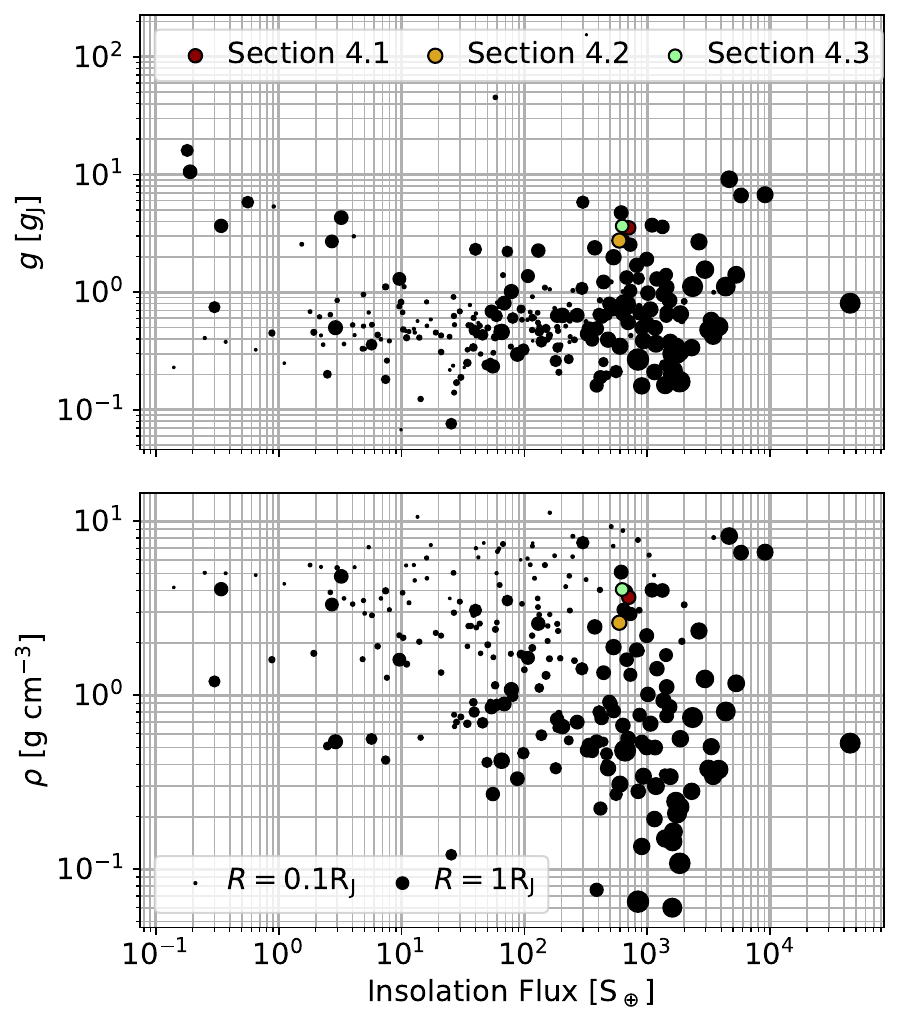}
    \caption{Grid of all exoplanets with published masses, radii, and insolation fluxes. The relative point size corresponds to planet radius. Top: surface gravity $g$ versus insolation flux $S$. Bottom: planet density $\rho$ versus insolation flux $S$. The three planetary models for KELT-24~b are shown here, and it is, regardless of choice, among the most dense, high-gravity, highly irradiated exoplanets. All data are sourced from the NASA exoplanet archive.}
    \label{fig:irr_dens_grav}
\end{figure}

KELT-24~b also presents a unique opportunity to study the effects that high surface gravity can have on atmospheres. Few planets experience as much stellar irradiance and are as dense KELT-24~b. Figure \ref{fig:irr_dens_grav} highlights KELT-24~b's placement in the landscape of well-known transiting planets. Among giant $\left(R_\mathrm{p}>1~\mathrm{R_J}\right)$ hot Jupiters, KELT-24~b is among the most dense and highly irradiated ($S\sim800~\mathrm{S_\oplus}$). As a result, future detailed atmospheric studies of KELT-24~b from space-based missions like JWST will provide a greater understanding of the formation and evolution of giant, high surface gravity hot Jupiters. The large mass of KELT-24~b occupies an important space for these types of studies that have preferentially targeted lower mass hot Jupiters and brown dwarfs.

\begin{figure}[h]
    \centering
    \includegraphics[width=\linewidth]{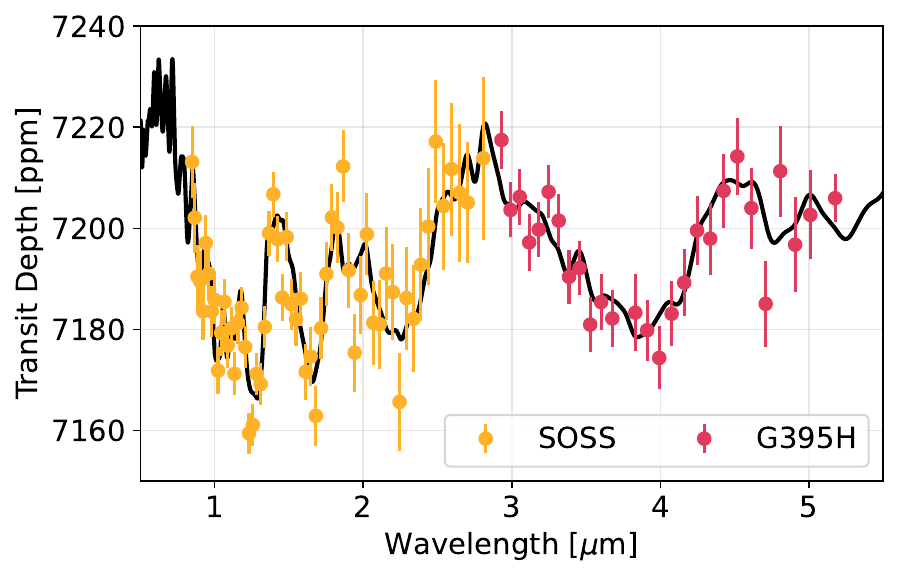}
    \caption{Simulated NIRISS SOSS (orange) and NIRSpec G395H (red) transmission spectra of KELT-24~b using four transit observations with each of the instruments generated using Pandexo \citep{2017PASP..129f4501B}. The underlying model assumes a cloud-free, solar-metallicity atmosphere, which should be broadly representative given KELT-24~b's mass and equilibrium temperature, and was created using Exo-Transmit \citep{2017PASP..129d4402K}. These observations would be able to strongly constrain the overall metallicity in the atmosphere, as well as the specific atmospheric abundances of water, methane, carbon dioxide, and carbon monoxide. In principle, these constraints would also measure KELT-24~b's atmospheric C/O ratio, which could place interesting constraints on the planet's formation pathway.}
    \label{fig:transmission_spec_sim}
\end{figure}

Element abundance ratios are particularly useful in understanding planet formation conditions \citep{2014prpl.conf..739M}, which is a key point of interest for well-aligned hot Jupiters like KELT-24~b. Giant planets on short orbits are generally believed to have undergone significant migration after forming. High-precision observations of KELT-24~b’s atmosphere will constrain abundance ratios, such as carbon-to-oxygen (C/O), to compare to predictions to yield insight into where the planet originally formed in the system’s disk \citep[e.g.,][]{2011ApJ...743L..16O, 2021JGRE..12606629F}. In Figure \ref{fig:transmission_spec_sim}, we simulate transmission spectra of the atmosphere of KELT-24~b using JWST's NIRISS Single Object Slitless Spectroscopy (SOSS; \citealt{2012SPIE.8442E..2RD}) and NIRSpec G395H \citep{2022A&A...661A..80J}, demonstrating the absorption features that are capable of being detected. We note that differences in the relative strength of these features across the wavelength range are negligible between our three planet models presented here, and so we only show the resulting simulated spectra assuming the best-fit parameters from Section \ref{subsec:no_sed}, which excludes the SED. In any case, KELT-24~b is an excellent candidate for future atmospheric studies from next-generation space missions.

R2019 and H2019 originally determined KELT-24~b's best-fit eccentricity to be $e=0.077^{+0.024}_{-0.025}$ and $e=0.050^{+0.020}_{-0.017}$, respectively, but as we have discussed, planets on true circular orbits may be found to have nonzero eccentricities as a result of statistical biases or too few observations. Indeed, our updated analysis reduce the best-fit eccentricity of KELT-24~b to $e=0.0319^{+0.0079}_{-0.0074}$, a higher-confidence result that KELT-24~b may be on a near-circular orbit. Note also that each of our three published models prefer a lower eccentricity than those reported by R2019 and H2019. Given the estimated youth of the system ($\mathrm{Age}=1.98^{+0.95}_{-0.79}$~Gyr), the identity of KELT-24~b's eccentricity is of particular interest, as it may offer insight into the dynamical history of the KELT-24 system.

\begin{figure}[h]
    \centering
    \includegraphics[width=\linewidth]{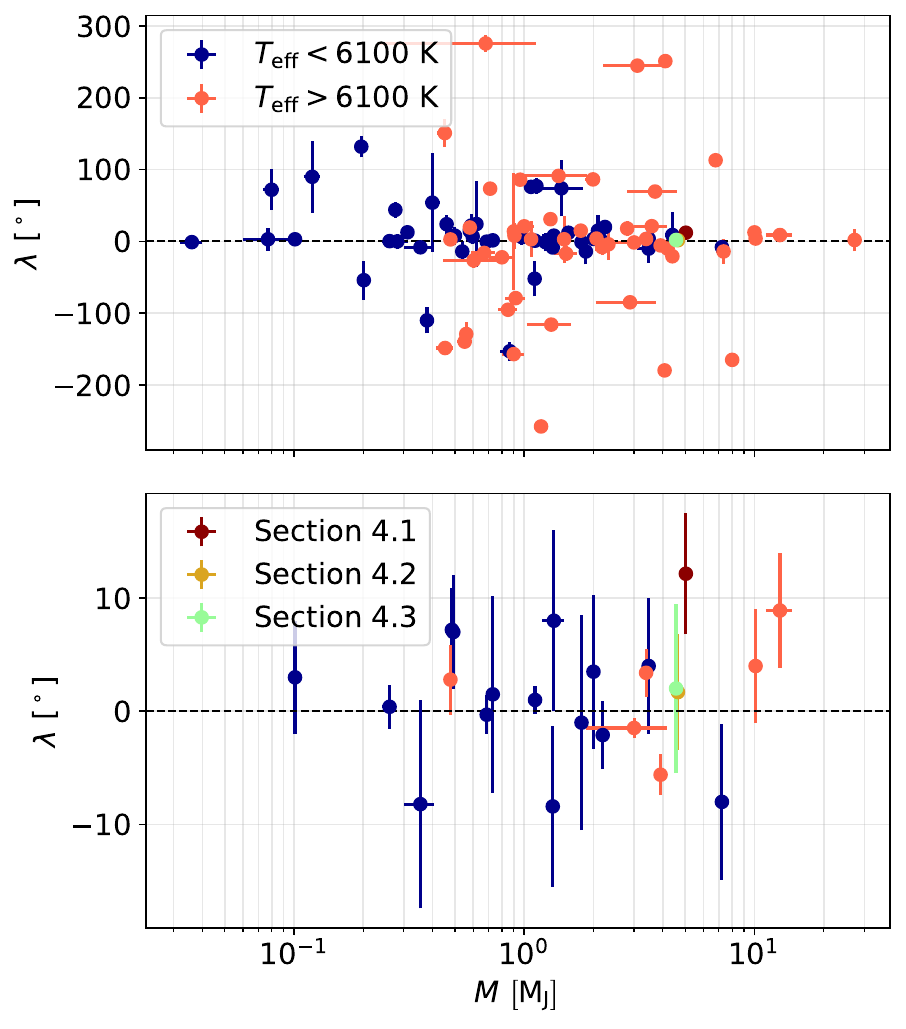}
    \caption{Top panel: distribution of exoplanets with known projected obliquities, divided at a stellar temperature of $T_\mathrm{eff}=6100~\mathrm{K}$ per the Kraft break \citep{1967ApJ...150..551K}. We note that all three of our stellar models derived here estimate an effective stellar temperature for KELT-24 well above 6100~K. Bottom panel: the same, but narrowed in on projected obliquities for well-aligned $\left(\vert\lambda\vert<10^\circ\mathrm{,~}\vert\sigma_\lambda\vert<10^\circ\right)$ planets. KELT-24~b is observed to be among the most massive planet with an orbit that is well-aligned to its host star.}
    \label{fig:rel_obliquities}
\end{figure}

Our updated constraints on KELT-24~b's spin-orbit angle ($\lambda=1.4^{+4.9}_{-4.7}~^\circ$) further validate the hot Jupiter as a well-aligned companion to its host star. KELT-24~b's alignment information represents a valuable data point to an otherwise sparse set of hot Jupiters with known relative obliquities. KELT-24~b is compared to other planets with known relative obliquities in Figure \ref{fig:rel_obliquities}. The KELT-24 planetary system is therefore a useful laboratory for understanding planet formation theories. 

\begin{figure}[h]
    \centering
    \includegraphics[width=\linewidth]{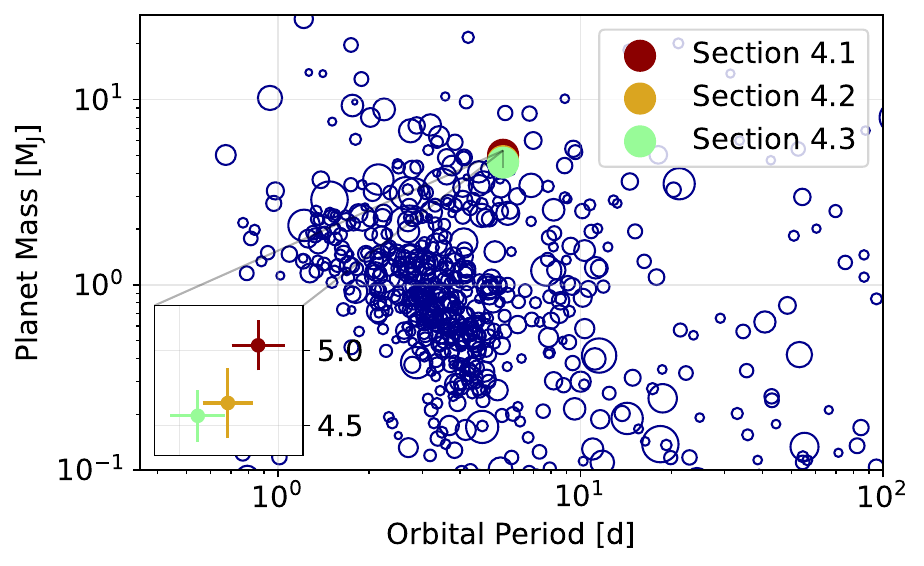}
    \caption{Collection of all transiting planets with measured masses, where the point size corresponds to the host star's brightness. KELT-24 is in all cases among the brightest stars with a massive, transiting hot Jupiter.}
    \label{fig:mass_per}
\end{figure}

\section{Conclusion} \label{sec:conclusion}

We present three separate global analyses of the KELT-24 system, each with a different approach for deriving the stellar template. These three trials were meant to identify the source of and bring resolution to the observed disparity between the original discovery papers (R2019 \& H2019). To aid in these efforts, we collate timeseries photometry from KELT-FUN and TESS, as well as RVs from TRES, SONG, and MINERVA to refine the orbit of KELT-24~b, a neighboring ($d=96.9$~pc) hot Jupiter on a 5.55~d period orbiting a bright ($V=8.3$), young F-type star. KELT-24~b is an excellent candidate for atmospheric characterization. Our updated timing ephemerides make follow-up observations of KELT-24~b more tenable.

When compared to the original discovery by R2019, the global fit described in Section \ref{sec:global_fit} refines the precision of KELT-24~b's orbital period by more than a factor of 10 to $\sim45$~ms. With the inclusion of $>200$ new RVs from MINERVA, as well as eight new sectors of TESS photometry, our favored approach to resolving the KELT-24 system finds $M_\mathrm{p}=4.59\pm0.18~\mathrm{M_J}$ and $R_\mathrm{p}=1.134^{+0.027}_{-0.026}~\mathrm{R_J}$ for KELT-24~b. This updates the planet density to $3.90^{+0.19}_{-0.20}~\mathrm{g~cm^{-3}}$, which is in the 95th percentile for known hot Jupiters. Our additional observations constrain the orbital eccentricity ($e=0.0319^{+0.0079}_{-0.0074}$) to be lower (e.g. a more circular orbit) than previously reported by either R2019 or H2019. This, combined with the planet's best-fit age ($1.98^{+0.95}_{-0.79}~\mathrm{Gyr}$) and well-aligned orbit ($\lambda=1.4^{+4.9}_{-4.7}~^\circ$), suggests a quick, yet calm migration of KELT-24~b via methods like in situ formation.

We have rigorously explored the derivation of KELT-24's stellar parameters. The independent analyses published by R2019 and H2019, while effectively confirming the existence of a transiting hot Jupiter orbiting KELT-24, posed serious questions regarding the system's stellar parameters that warranted further study. The noticeable difference between posteriors between the independent analyses presented in Section \ref{sec:global_fit} affirm some the significant disagreements that can be seen, even across independently self-consistent global fits. This confusion remains an open question. It is imperative to note that none of our analyses shown here are necessarily absolute. Rather, each should be consistent with another, and the fact that that it is not the case is likely indicative of systematic errors. In order to assess KELT-24~b in the landscape of known exoplanets, we posit the SED to be the most likely culprit for the observed discrepancies in the various stellar parameters, though additional imaging and high-resolution spectroscopy will be needed to further rule out fainter sources that may be the exact cause for potential contamination. Specifically, we recommend an isolated spectroscopic observation of KELT-24B to We therefore adopt the system parameters derived in Section \ref{subsec:no_sed}. Our case study of KELT-24 promotes deeper investigations of stellar fits for future global analyses of exoplanet systems.

~
~
~
~
~
~

\acknowledgements 

We thank the anonymous referee for their thoughtful comments and careful review of our manuscript that contributed to the improvement of the quality and clarity of our research. Both the Fred Lawrence Whipple Observatory, which hosts MINERVA and TRES, as well as Winer Observatory, which was the site of KELT-North, are built upon lands ancestral to people from the Tohono O'odham Nation, Ak-Chin Indian Community, Hia-Ced O'odham, and Hohokam. The two AO images used in this analysis were taken at the W. M. Keck Observatory on Maunakea, a dormant volcano that has been sacred to native Hawaiians for centuries.

This research has made use of the SIMBAD database and the VizieR catalogue access tool, both operated at CDS, Strasbourg, France. Additionally, our work has made use of the NASA Exoplanet Archive, which is operated by the California Institute of Technology, under contract with the National Aeronautics and Space Administration under the Exoplanet Exploration Program. This work has made use of data from the European Space Agency (ESA) mission {\it Gaia} (\url{https://www.cosmos.esa.int/gaia}), processed by the {\it Gaia} Data Processing and Analysis Consortium (DPAC, \url{https://www.cosmos.esa.int/web/gaia/dpac/consortium}). Funding for the DPAC has been provided by national institutions, in particular the institutions participating in the {\it Gaia} Multilateral Agreement.

Work by J.D.E. was funded NASA ADAP 80NSSC19K1014. MINERVA is a collaboration among the Harvard-Smithsonian Center for Astrophysics, The Pennsylvania State University, the University of Montana, the University of Southern Queensland, University of Pennsylvania, George Mason University, and the University of New South Wales. It is made possible by generous contributions from its collaborating institutions and Mt. Cuba Astronomical Foundation, The David \& Lucile Packard Foundation, National Aeronautics and Space Administration (EPSCOR grant NNX13AM97A, XRP 80NSSC22K0233), the Australian Research Council (LIEF grant LE140100050), and the National Science Foundation (grants 1516242, 1608203, and 2007811). Funding for the TESS mission is provided by NASA’s Science Mission Directorate. This paper includes data collected by the TESS mission, which are publicly available from the Mikulski Archive for Space Telescopes (MAST). We acknowledge the use of public TESS data from pipelines at the TESS Science Office and at the TESS Science Processing Operations Center. Resources supporting this work were provided by the NASA High-End Computing (HEC) Program through the NASA Advanced Supercomputing (NAS) Division at Ames Research Center for the production of the SPOC data products.

\bibliography{bib_master}{}
\bibliographystyle{aasjournal}



\end{document}